\documentclass[journal,10pt]{IEEEtran}

\usepackage{cite}
\usepackage{amsmath,amssymb,amsfonts}
\usepackage{algorithmic}
\usepackage{graphicx}
\usepackage{textcomp}
\usepackage{xcolor}
\usepackage{mathtools}
\usepackage{cite}
\usepackage{multirow}
\usepackage{bm}
\usepackage{adjustbox,bm}
\usepackage{tikz}
\usepackage[justification=centering]{caption}
\usetikzlibrary{fit,tikzmark,arrows.meta,calc}
\usepackage[caption=false,font=footnotesize,labelfont=sf,textfont=sf]{subfig}

\def\BibTeX{{\rm B\kern-.05em{\sc i\kern-.025em b}\kern-.08em
    T\kern-.1667em\lower.7ex\hbox{E}\kern-.125emX}}
\begin{document}

\title{MIMO Zak-OTFS: Channel Estimation, Detection, and Throughput Analysis
}

\author{Faraz~Barati$^{*}$,~Rahul~Kumar~Jaiswal$^{*}$,~Saif~Khan~Mohammed,~Ronny~Hadani,~and~Jeffrey~G.~Andrews%
\thanks{$^{*}$F. Barati and R. K. Jaiswal contributed equally to this work.}
\thanks{F. Barati, R. Hadani, and J. G. Andrews are with The University of Texas at Austin, Austin, TX, USA. 
F. Barati and J. G. Andrews are with the Chandra Department of Electrical and Computer Engineering, 
and R. Hadani is with the Department of Mathematics. All are with 6G@UT and the Wireless Networking and Communication Group (WNCG). Ronny Hadani is also with the Cohere Technologies
Inc., San Jose, CA, USA. (email: faraz.barati@utexas.edu, jandrews@ece.utexas.edu, hadani@math.utexas.edu).}
\thanks{R. K. Jaiswal and S. K. Mohammed are with the Department of Electrical Engineering,
Indian Institute of Technology Delhi, New Delhi, India (email: Rahul.Kumar.Jaiswal@ee.iitd.ac.in,
saifkmohammed@gmail.com). S. K. Mohammed is currently on extra-ordinary leave from the
Indian Institute of Technology Delhi and is with Cohere Technologies Inc., CA, USA.}
\thanks{Manuscript last updated: \today.}
}

\maketitle

\begin{abstract}
Zak-Orthogonal Time Frequency Space (Zak-OTFS) modulation has demonstrated substantial performance gains over cyclic-prefix orthogonal frequency-division multiplexing (CP-OFDM) in highly time and frequency selective channels. In this paper, we extend Zak-OTFS to a multiple-input multiple-output (MIMO) framework. We first derive a complete system model for MIMO Zak-OTFS based directly on the physical multipath channel; ours is the first work to do so.  We then propose an efficient channel estimation method using structured pilot placement in the delay–Doppler (DD) domain. The proposed approach is evaluated under the standardized CDL-C channel model, demonstrating that the advantages of Zak-OTFS observed in SISO scenarios extend to MIMO systems, particularly its robustness to Doppler and inter-carrier interference (ICI). We identify a fundamental crossover behavior: CP-OFDM performs slightly better at low SNR and low Doppler, while Zak-OTFS excels at higher SNR or under severe Doppler dispersion. Furthermore, we show that the crossover points for SNR and Doppler shift inversely to each other.  We also observe that Zak-OTFS -- particularly with MIMO -- exhibits increased sensitivity to high values of pilot-to-data power ratio (PDR), but has a similar optimal PDR as CP-OFDM.

\end{abstract}


\section{Introduction}
The evolution to sixth-generation (6G) wireless systems is expected to support highly dynamic environments with carrier frequencies extending into the upper mid band ranges, ultra-dense antenna arrays, and mobility profiles that induce rapid channel variations, as encountered in emerging applications such as Vehicle-to-Everything (V2X), unmanned aerial vehicles (UAVs), low Earth orbit (LEO) satellite Direct-to-Cell (D2C) communications, and Joint Communication and Sensing (JCAS) systems \cite{6g,6g_takes_shape,noor_v2x,wang_v2x,d2d}.
Such conditions result in \emph{doubly dispersive} propagation, where both delay and Doppler spreads jointly impact link reliability and channel coherence. Achieving robust and efficient transmission in these channels motivates the exploration of alternative modulation and estimation frameworks that go beyond the time–frequency separability assumption of cyclic-prefix orthogonal frequency division multiplexing (CP-OFDM). In such environments, CP-OFDM suffers from inter-carrier interference (ICI) and inter-symbol interference (ISI), leading to a loss of subcarrier orthogonality, inaccurate channel estimation, and reduced spectral efficiency as the channel can no longer be assumed constant over an OFDM symbol interval. Although CP-OFDM can mitigate fading, ISI, and Doppler through numerology adaptation (as in 5G) or other signal processing methods~\cite{schnider}, the resulting trade-offs between Doppler robustness, delay-spread tolerance, and pilot overhead limit its effectiveness in highly doubly dispersive channels. These limitations motivate the exploration of modulation schemes that can capture the inherent coupling between time and frequency dispersions. Delay-Doppler (DD) domain representations provide a natural framework for modeling doubly selective channels, and orthogonal time frequency space (OTFS) modulation directly represents and equalizes channel effects in the DD domain~\cite{otfs,BestReadingsOTFS,Hadani2018OTFSarxiv}. The Zak-transform-based formulation (Zak-OTFS) further refines this approach by employing quasi-periodic DD-domain pulses, referred to as \emph{pulsones}~\cite{math_foundation,otfs_book}, revealing the underlying DD lattice structure, unlike the multi-carrier OTFS (MC-OTFS) implementation~\cite{Zak_vs_ofdm,mcvsZak,Zak_ldpc}. 

Despite the growing interest in OTFS modulation schemes, there has been limited work on the integration of Zak-OTFS with multiple-input multiple-output (MIMO) systems, which are foundational for increasing spectral efficiency in current wireless systems~\cite{5mimo,5G,mmimo}. As we move towards 6G, MIMO will remain a key architectural element, supporting increasingly difficult requirements on data rates, reliability, and coverage~\cite{6g_takes_shape,6G1,6G2}. Consequently, for Zak-OTFS (or any other waveform) to be adopted in future wireless systems, it must be able to operate synergistically with MIMO. However, extending Zak-OTFS to the MIMO setting introduces new challenges in channel modeling, signal representation, and receiver design. While SISO Zak-OTFS is described by a scalar DD channel kernel and its associated twisted-convolution structure~\cite{otfs_book}, a MIMO formulation must jointly capture spatial, delay, and Doppler interactions. A naive extension based on independently stacking SISO channel matrices neither fully reflects the underlying propagation physics nor guarantees preservation of the structural properties exploited in SISO Zak-OTFS. Consequently, a key challenge is to develop a MIMO DD channel representation that incorporates spatial signatures while retaining the mathematical structure required for efficient signal processing.

MIMO Zak-OTFS also introduces new challenges in channel estimation and pilot design. Although SISO Zak-OTFS supports efficient embedded pilot schemes for DD domain channel acquisition, extending these methods to MIMO is complicated by the simultaneous transmission of multiple pilot signals and the resulting interference. Furthermore, straightforward pilot separation across transmit antennas can lead to pilot overhead that scales unfavorably with the number of antennas. Consequently, MIMO Zak-OTFS requires pilot and estimation strategies that jointly address spatial channel separation, estimation accuracy, and spectral efficiency. In this paper, we develop a comprehensive MIMO Zak-OTFS framework that addresses these challenges and investigate its performance in doubly selective channels.

\subsection{Background and Related Works}
The early Zak OTFS work~\cite{mcvsZak} introduced the \emph{crystallization condition}, under which the Zak-OTFS input-output (I/O) relation becomes predictable and effectively eliminates fading. Under this condition, the received DD domain signal is related to the transmitted DD domain information signal through a two-dimensional twisted convolution with an effective DD domain channel filter. Furthermore, when the Zak-OTFS delay and Doppler period parameters exceed the corresponding channel spreads, this effective channel filter can be directly obtained from the received response to a single pulsone waveform. One of the key trade-offs of this modulation scheme is that it provides a simplified and structured representation of the channel while inherently accounting for both ICI and fading. However, this comes at the cost of increased complexity of joint equalization of all information symbols.

To address this challenge, \cite{comp1,comp2} propose low-complexity Zak-OTFS equalization by developing a frequency-domain formulation that is unitarily equivalent to delay–Doppler processing. By exploiting the banded structure of the resulting channel matrix, they employ conjugate-gradient methods to achieve near-MMSE performance with linear complexity.
For practical channel estimation, the authors in \cite{predictability} propose a same-frame pilot--data structure, where distinct regions are allocated to pilots, data, and guard intervals for simultaneous channel sensing and communication. However, similar to OFDM, this approach suffers from a high peak-to-average power ratio (PAPR).

To address this issue, the authors in \cite{Zak_io} introduce \emph{spread pulsones}, generated via discrete chirp filtering, which effectively reduce the PAPR while enabling joint communication and sensing within a single Zak-OTFS subframe. Their results showed that orthogonal pilot and data pulsones on rotated lattices can coexist incoherently, allowing simultaneous channel estimation and data transmission without explicit sharing of DD resources. Extending this framework,~\cite{predictability} refined pilot and pulse-shaping design using Gaussian DD-domain filters and structured pilot–guard–data subframes, reducing aliasing and improving BER over conventional sinc and root-raised-cosine filters. Complementary works~\cite{Gginc,closedform} derived closed-form DD-domain channel models for various pulse shapes, showing that Gaussian filters outperform other filters in terms of channel estimation via low sidelobes, while Gauss–Sinc filters achieve better BER through a balanced main-lobe–sidelobe tradeoff.

An important step towards practical realization of Zak-OTFS is time-frequency based windowing~\cite{predictability,Zak_implementation,optimal_receiver}. Gopalam \emph{et al.}~\cite{Zak_implementation,optimal_receiver} identified two classes of DD filters (Type-1 and Type-2), realizable via interpolation filtering and precoded-OFDM structures, respectively, with Type-2 offering higher spectral efficiency. In subsequent work, optimal Zak-OTFS receiver architectures were developed by interpreting demodulation as correlation-based processing governed by the TC filter. This led to matched TC filters that maximize SNR, extensions to doubly dispersive channels, and both radar-inspired matched-filter and practical windowing-based implementations, which were shown to be equivalent under sufficient support and analogous to rake receivers in the delay–Doppler domain.
An implementation of Zak-OTFS over exisiting 4G and 5G CP-OFDM based modem was proposed in~\cite{zak_over_ofdm}. This Zak-OTFS-over-CP-OFDM architecture embeds Zak-OTFS as a low-complexity precoder and post-processor within existing OFDM hardware. This design achieved substantial spectral efficiency gains under high mobility while maintaining backward compatibility with legacy infrastructure.

An important next step is to explore the MIMO extension of Zak-OTFS, leveraging spatial multiplexing gains to enhance link reliability and spectral efficiency. The authors in~\cite{mcmimo} investigated MIMO extensions of MC-OTFS and demonstrated its advantages over SISO-MC-OTFS through analytical modeling and simulations. More recently,~\cite{Zakmimo} study $2\times2$ MIMO Zak-OTFS with separate frames for channel estimation and data (i.e., pilot and data are not transmitted together), which is not realistic. In~\cite{spreadmimo}, MIMO Zak-OTFS with spread pilot pulsones
overlaid on data pulsones was proposed, but suffered from poor performance and is able to only decode BPSK modulated information symbols due to the interference between data pulsones and the spread pilot pulsones which limits its ability to achieve high throughput. Also~\cite{Zakmimo,spreadmimo} do not provide a comprehensive study of the achievable spectral efficiency and its dependence on the pilot arrangement.

\subsection{Contributions}
We propose a novel pilot placement strategy that enables the extension of MIMO to Zak-OTFS without incurring a linear increase in pilot overhead (in CP-OFDM, pilot overhead increases
linearly with increase in number of transmit antennas). Specifically, pilots are staggered across both the delay and Doppler dimensions, effectively mitigating interference between the channel response to pilots sent from different transmit antennas, while allowing all pilots to be embedded along with data within the same frame. The main contributions of this paper are as follows:
\begin{itemize}
    \item We derive the DD I/O relation for MIMO Zak-OTFS based on the physical multi-path model and establish conditions under which \emph{crystallization} holds for each pair of transmit and receive antennas.
    \item We propose an efficient pilot placement strategy that maintains interference-free reception of pilots transmitted from different transmit antennas in DD domain, while reusing the same pilot–guard–data subframe structure used in single-antenna Zak-OTFS.
    \item We evaluate performance in terms of normalized mean square error (NMSE), bit error rate (BER), and achieved spectral efficiency under fractional delay--Doppler spreading using the standardized CDL-C channel model. We compare the proposed MIMO Zak-OTFS with CP-OFDM under both perfect and imperfect channel state information (CSI), and a wide variety of Doppler shifts and SNRs.
\end{itemize}

We demonstrate that CP-OFDM faces a fundamental limitation in highly doubly dispersive channels, where it cannot simultaneously suppress both ICI and ISI without sacrificing throughput resources. In contrast, Zak-OTFS explicitly models and equalizes both impairments in the DD domain. Nevertheless, Zak-OTFS channel estimation is limited by AWGN, resulting in a characteristic crossover behavior of performance comparison with CP-OFDM: CP-OFDM outperforms Zak-OTFS at low SNR or low Doppler, while Zak-OTFS achieves superior performance at higher SNRs or under moderate to severe Doppler conditions, where OFDM becomes ICI-limited. Furthermore, the proposed pilot placement strategy improves spectral efficiency without incurring a proportional increase in pilot overhead. 
Overall, these results highlight MIMO Zak-OTFS as a compelling waveform candidate for reliable high-rate communication in high mobility scenarios.

\subsection{Organization $\And$ Notation}
The remainder of this paper is organized as follows. Section II presents the SISO and MIMO system models for CP-OFDM and Zak-OTFS. Section III introduces the proposed pilot placement and channel estimation strategy. Section IV provides performance comparisons under various channel conditions, and Section V concludes the paper with directions for future work. \\
\textit{Notation}: $\mathbf{A}$ is a matrix and $\mathbf{a}$ is a vector. A continuous function is denoted as $A(t)$ while the discretized one is denoted as $A[t]$. $\|\mathbf{A}\|_F^2$ denotes the Frobenius norm of
$\mathbf{A}$. $N_r$, $N_t$, $\nu$ and $\tau$ denote the number of receive antennas, number of transmit antennas, Doppler shift value and Delay value respectively.

\section{System Model}
In this section, we first present the MIMO CP-OFDM system model as the baseline reference. We then introduce the Zak-OTFS system model for the SISO case, followed by its extension to the MIMO setting.
A doubly-spread wireless channel exhibits both time and frequency selectivity. 
For a transmitted time-domain signal $s_{\mathrm{td}}(t)$, the corresponding received signal $r_{\mathrm{td}}(t)$ 
after propagation through such a channel can be expressed as in \cite{doublyspread}:
\begin{equation}
    r_{\mathrm{td}}(t) = 
    \iint h_{\mathrm{phy}}(\tau,\nu)\, s_{\mathrm{td}}(t-\tau)\,
    e^{j2\pi\nu (t-\tau)} \, d\tau\, d\nu + n_{\mathrm{td}}(t),
\end{equation}
where $h_{\mathrm{phy}}(\tau,\nu)$ denotes the physical delay–Doppler domain channel response and 
$n_{\mathrm{td}}(t)$ is additive white Gaussian noise (AWGN). 
For a multi-path environment with $L$ discrete propagation paths, 
$h_{\mathrm{phy}}(\tau,\nu)$ can be modeled as
\begin{equation}
    h_{\mathrm{phy}}(\tau,\nu) 
    = \sum_{i=1}^{L} h_i\, \delta(\tau - \tau_i)\, \delta(\nu - \nu_i),
\end{equation}
where $h_i$, $\tau_i$, and $\nu_i$ denote the complex gain, delay, 
and Doppler shift associated with the $i$-th path, respectively.

\subsection{MIMO CP-OFDM Baseline}
For reference, we consider a conventional CP-OFDM system with $K$ subcarriers and subcarrier spacing $\Delta f$.
The transmitted baseband signal from the $u$-th antenna is
\begin{equation}
    s_{\mathrm{td}}^{(u)}(t) =
    \sum_{i=0}^{K-1} S_u[i]\,
    e^{j2\pi i \Delta f t} W_2(t),
\end{equation}
where $S_u[i]$ denotes the frequency domain data symbol transmitted over the $i$-th subcarrier from the $u$-th transmit antenna. Furthermore,
\begin{equation}
    W_2(t) =
    \begin{cases}
      \dfrac{1}{\sqrt{T}}, & -T_{\mathrm{cp}} \le t < T, \\[3pt]
      0, & \text{otherwise},
    \end{cases}
\end{equation}
and $T = 1/\Delta f$ and $T_{\mathrm{cp}}$ denotes the cyclic prefix duration.
The composite received signal at the $v$-th antenna is given by
\begin{align}
r_{\mathrm{td}}^{(v)}(t)
&= \sum_{u=0}^{N_t-1} 
   \iint h_{\mathrm{phy}}^{(v,u)}(\tau,\nu)\,
        s_{\mathrm{td}}^{(u)}(t-\tau)\,
        e^{j2\pi \nu (t-\tau)}\, d\tau\, d\nu \notag\\[-2pt]
&\quad + n_{\mathrm{td}}^{(v)}(t),
\label{eq:rtd_rx}
\end{align}

where $h_{\mathrm{phy}}^{(v,u)}(\tau,\nu)$ is the doubly selective DD channel between transmit antenna $u$ and receive antenna $v$. Using the DFT-domain representation, the overall MIMO frequency-domain channel can be expressed as
\begin{equation}
    \mathbf{H} =
    \begin{pmatrix}
        \mathbf{H}[0,0] & \mathbf{H}[0,1] & \cdots & \mathbf{H}[0,K-1] \\
        \mathbf{H}[1,0] & \mathbf{H}[1,1] & \cdots & \mathbf{H}[1,K-1] \\
        \vdots & \vdots & \ddots & \vdots \\
        \mathbf{H}[K-1,0] & \mathbf{H}[K-1,1] & \cdots & \mathbf{H}[K-1,K-1]
    \end{pmatrix},
\end{equation}
where $\mathbf{H}[i,k]\in\mathbb{C}^{N_r\times N_t}$ denotes the MIMO frequency-domain channel matrix describing the coupling from the $k$th subcarrier to the $i$th subcarrier. After CP removal and DFT demodulation, the received signal on the $i$th subcarrier can be written as:

\begin{equation}
\begin{aligned}
\mathbf{y}[i]
&=
\mathbf{H}[i,i]\mathbf{s}[i]
+
\sum_{k\neq i}\mathbf{H}[i,k]\mathbf{s}[k]
+
\mathbf{n}[i], \\
&\hspace{1cm}\qquad i,k \in \{0,1,\ldots,K-1\}.
\end{aligned}
\end{equation}
where $\mathbf{y}[i]\in\mathbb{C}^{N_r\times1}$ and
$\mathbf{s}[i]\in\mathbb{C}^{N_t\times1}$ denote the received and transmitted signal vectors on the $i$th subcarrier, respectively. The first term represents the desired signal contribution, while the second term ICI arising from Doppler-induced loss of subcarrier orthogonality. In the absence of ICI, $\mathbf{H}[i,k]=\mathbf{0}$ for all $i\neq k$, and the channel matrix becomes block diagonal.

Per-subcarrier spatial equalization is typically performed as
$\hat{\mathbf{s}}[i]=\mathbf{W}[i]\mathbf{y}[i]$,
where $\mathbf{W}[i]$ denotes an MMSE equalizer. Under high Doppler spread, the frequency-domain channel matrix is no longer block diagonal after CP removal and FFT processing, resulting in significant ICI and degraded OFDM performance. This limitation motivates the Zak-OTFS framework.

\subsection{SISO Zak OTFS}
\begin{figure*}[t]
    \centering
    \includegraphics[width=0.95\textwidth]{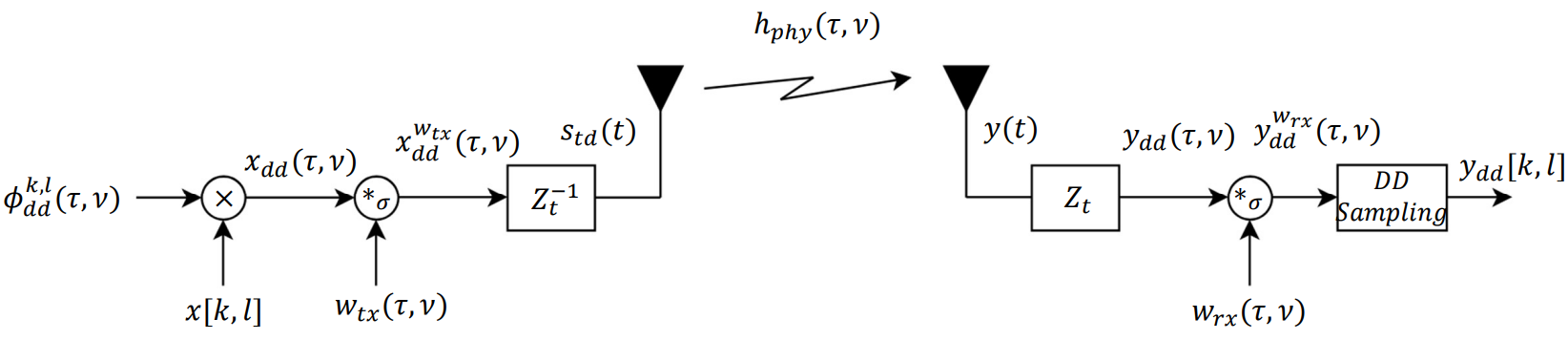}
    \caption{SISO Zak-OTFS System Model}
    \label{fig:siso_model}
\end{figure*}

In Fig~\ref{fig:siso_model}, we illustrate the SISO system model.
In Zak-OTFS modulation, information is carried by quasi-periodic pulses in the DD domain. A DD domain quasi-periodic pulse has period $\tau_p$ and $\nu_p = 1/\tau_p$ along the delay and Doppler axis respectively \cite{Zak_io}. In Zak-OTFS modulation, information carrying pulses are located at points of the information lattice:
\begin{equation}
    \Lambda_{dd} = \{(k\tau_p/M,l\nu_p/N)|k,l \in \mathbb{Z} \}.
\end{equation}
The time-domain realization of a DD domain pulse is a pulse train modulated by a tone, and is therefore referred to as pulsone \cite{otfs_book,mcvsZak}. A DD pulse located at  $\left(\tfrac{k\tau_p}{M}, \tfrac{l\nu_p}{N}\right)$ with bandwidth $B = \tfrac{M}{\tau_p}$ and duration 
$T = \tfrac{N}{\nu_p}$ can be expressed as:

\begin{equation}
\begin{aligned}
\phi_{dd}^{k,l}(\tau,\nu)=
\sum_{n \in \mathbb{Z}} \sum_{m \in \mathbb{Z}}
e^{j 2\pi \frac{nl}{N}}
&\delta\!\left(\tau-n\tau_{p}-\frac{k\tau_{p}}{M}\right)  \\
&\delta\!\left(\nu-m\nu_{p}-\frac{l\nu_{p}}{N}\right).
\end{aligned}
\end{equation}

Here, $M$ and $N$ are positive integers, such that the time duration and bandwidth of the time-domain realization of a DD pulse is $T = N \tau_p$ and $B = M \nu_p$ respectively. The transmission begins with a two-dimensional array of information symbols $x[k,l]$, where $k = 0, \ldots, M-1$ and $l = 0, \ldots, N-1$. The DD domain information signal that carries all $M\times N$ information symbols is obtained as:
\begin{equation}
    x_{dd}(\tau,\nu) = \sum_{k = 0}^{M-1}\sum_{l = 0}^{N-1}x[k,l]\phi_{dd}^{k,l}(\tau,\nu).
\end{equation}
Equivalently, $x_{dd}(\tau,\nu)$ can be acquired by lifting the the discrete information signal $x_{dd}[\cdot,\cdot]$ to the information lattice:
\begin{equation}\label{xdd_kl}
x_{dd}[k,\, l] \;\triangleq\; e^{j \tfrac{2\pi}{N} nl}\, x[k\,\bmod\,M,l\, \bmod\,N].
\end{equation}

\begin{equation}
x_{dd}(\tau,\nu) \;\triangleq\; \sum_{k,l \in \mathbb{Z}}x_{dd}[k,l]\delta(\tau - k\tau_p/M)\delta(\nu - l\nu_p/N) .
\end{equation}
Note that $x_{dd}(\tau,\nu)$ is quasi-periodic as well:
\begin{equation}
x_{\mathrm{dd}}(\tau + n\tau_p, \nu + m\nu_p)
= e^{j2\pi n\nu\tau_p} \, x_{\mathrm{dd}}(\tau, \nu), 
\quad \forall n,m \in \mathbb{Z}.
\end{equation}
For guaranteeing the limitations of bandwidth and time duration the information signal is 
pulse-shaped with the transmit pulse-shaping filter $w_{tx}(\tau,\nu)$ by the means of twisted convolution. The twisted convolution between two DD functions $a(\tau,\nu)$ and $b(\tau,\nu)$ is defined as:
\[
\begin{aligned}
c(\tau,\nu)
&= a(\tau,\nu) \ast_\sigma b(\tau,\nu) \\[-2pt]
&\hspace{-0.5cm}= \!\!\int_{-\infty}^{\infty}\!\!\!\int_{-\infty}^{\infty}
a(\tau',\nu')\, b(\tau-\tau',\,\nu-\nu')\,
e^{j2\pi \nu'(\tau-\tau')} d\tau' d\nu'.
\end{aligned}
\]

Hence the DD domain pulse-shaped signal can be expressed as:
\begin{equation}
    x_{dd}^{w_{tx}}(\tau,\nu) \triangleq 
w_{tx}(\tau,\nu) \ast_\sigma x_{dd}(\tau,\nu).
\end{equation}
Note that twisted convolution preserves quasi-periodicity, i.e., $x_{dd}^{w_{tx}}(\tau,\nu)$ is also quasi-periodic. Finally the inverse Zak-transform $\left( \mathcal{Z}_t^{-1} \right)$ of the filtered DD domain signal $x_{dd}^{w_{tx}}(\tau,\nu)$ gives its time-domain realization which is the transmitted signal \cite{otfs_book}:
\begin{equation}
    s_{td}(t) = \mathcal{Z}_t^{-1}\!(x_{dd}^{w_{tx}}(\tau,\nu)) = \sqrt{\tau_p}\int_0^{\nu_p}x_{dd}^{w_{tx}}(t,\nu)d\nu .
\end{equation}
The DD representation of the received signal $r_{td}(t)$ is given by its Zak-transform $\left( \mathcal{Z}_t \right)$~\cite{otfs_book}:
\begin{equation}
    y_{dd}(\tau,\nu) = \mathcal{Z}_t\{r_{td}(t)\} = \sqrt{\tau_p} \sum_{n \in Z} r_{td}(\tau + n \tau_p) e^{- j 2 \pi n \nu \tau_p}.
\end{equation}
Matched filtering with the receive DD pulse $w_{rx}(\tau,\nu)$ results in
\begin{equation}
y_{dd}^{w_{rx}}(\tau,\nu) 
= w_{rx}(\tau,\nu) \ast_\sigma y_{dd}(\tau,\nu).    
\end{equation}
Since twisted convolution preserves quasi-periodicity, 
$y_{dd}^{w_{rx}}(\tau,\nu)$ is quasi-periodic. The equivalent DD domain additive noise is described as $n^{w_{rx}}_{dd}(\tau,\nu) = w_{rx}(\tau,\nu) \ast_\sigma n_{dd}(\tau,\nu)$, where $n_{dd}(\tau,\nu) = \mathcal{Z}_t\{n_{td}(t)\}$. Sampling $y_{dd}^{w_{rx}}(\tau,\nu)$ on the information lattice $\Lambda_{dd}$ gives:
\begin{equation}\label{ydd_kl}
    y_{dd}[k,l] = y_{dd}^{w_{rx}}\!\left(
\tau = \tfrac{k\tau_p}{M},\;
\nu = \tfrac{l\nu_p}{N}\right), \forall k,l \in \mathbb{Z}.
\end{equation}
From equations (\ref{xdd_kl})-(\ref{ydd_kl}), it follows that the SISO Zak-OTFS DD domain I/O relation between $x_{dd}[k,l]$ and $y_{dd}[k,l]$ is given by \cite{Zak_io}
\begin{align}
y_{dd}[k,l]
&= h_{\mathrm{eff}}[k,l] \ast_\sigma x_{dd}[k,l]
+ n_{dd}[k,l] \nonumber \\
&= \sum_{k',l'\in\mathbb{Z}}
h_{\mathrm{eff}}[k',l']\,x_{\mathrm{dd}}[k-k',\,l-l']
\notag \\
&\quad \times e^{j \tfrac{2\pi}{MN} l'(k-k')}
+ n_{\mathrm{dd}}[k,l].
\label{eq:Zak_io}
\end{align}
Here, the discrete noise $n_{dd}[k,l]$ and effective channel $h_{\mathrm{eff}}[k,l]$ are described as:
\begin{subequations}\label{eq:heff_definition}
\begin{align}
n_{\mathrm{dd}}[k,l]
&\triangleq n_{\mathrm{dd}}^{w_{\mathrm{rx}}}\!\left(\tau=\frac{k\tau_p}{M},\,\nu=\frac{l\nu_p}{N}\right), \label{eq:noisedd}\\
h_{\mathrm{eff}}[k,l]
&\triangleq h_{\mathrm{eff}}\!\left(\tau=\frac{k\tau_p}{M},\,\nu=\frac{l\nu_p}{N}\right), \label{eq:Ed}
\end{align}
\end{subequations}
where the effective continuous DD domain channel is given by
\begin{equation}
    h_{\mathrm{eff}}(\tau,\nu)= w_{\mathrm{rx}}(\tau,\nu) \ast_{\sigma} h_{\mathrm{phy}}(\tau,\nu) \ast_{\sigma} w_{\mathrm{tx}}(\tau,\nu), \label{eq:heff_cont}
\end{equation}

with $n_{dd}[k,l]$ being the discrete DD domain noise signal. For a given transmit pulse shaping filter $w_{tx}(\tau, \nu)$, the matched receive filter is given by \cite{optimal_receiver} 
\begin{equation}
    w_{rx}(\tau,\nu) = e^{j2\pi \nu \tau}\; w_{tx}^\ast(-\tau,-\nu).
\end{equation}

The quasi-periodicity of $y_{dd}[k,l]$ allows the detection of the transmitted
information symbols $x[k,l]$, where $k = 0, 1, \ldots, M-1$ and $l = 0, 1, \ldots, N-1$,
directly from the received symbols $y_{dd}[k,l]$ over the same index ranges.
This property leads to a matrix–vector representation of the input–output relationship,
in which the $MN \times 1$ vector containing all received symbols is expressed as the
product of an $MN \times MN$ effective channel matrix and the corresponding
$MN \times 1$ vector of transmitted information symbols:
\begin{equation}
    \mathbf{y} = \mathbf{H}\mathbf{x} + \mathbf{n}
    \label{sisomat}
\end{equation}
Where $\mathbf{y}\in \mathbb{C}^{MN \times 1}$, $\mathbf{x}\in \mathbb{C}^{MN \times 1}$, $\mathbf{n}\in \mathbb{C}^{MN \times 1}$, $\mathbf{H}\in \mathbb{C}^{MN \times MN}$.
\vspace{-0.5cm}
\subsection{MIMO Zak OTFS}
\begin{figure*}[!t]
    \centering
    \includegraphics[width=\textwidth,
                     height=0.445\textwidth,
                     keepaspectratio]{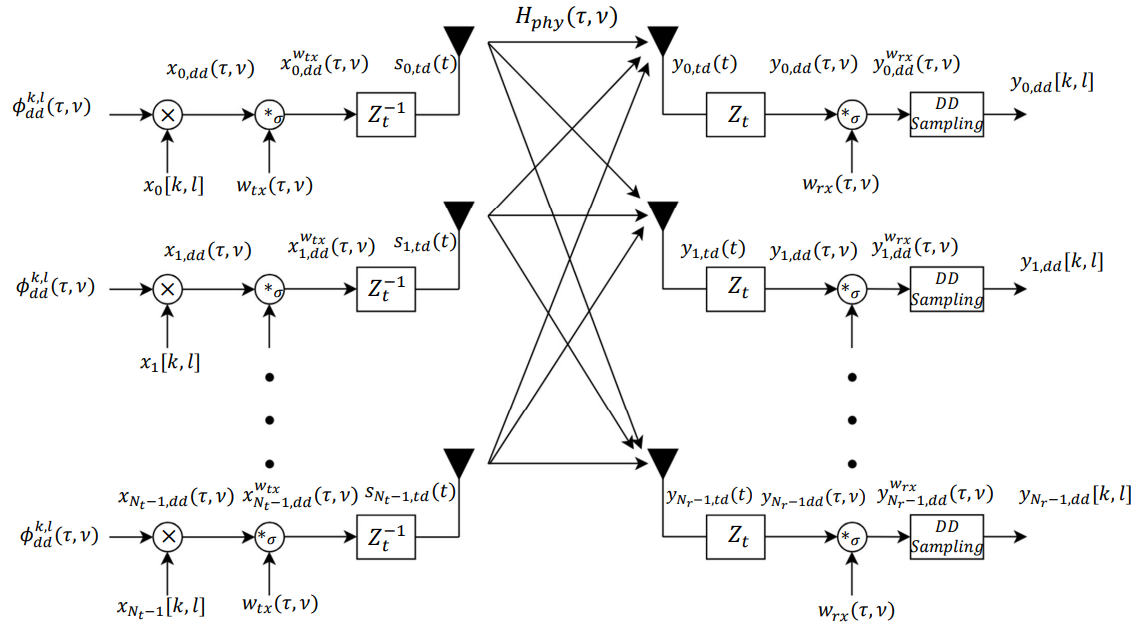}
    \caption{MIMO Zak-OTFS System Model.}
    \label{fig:mimo_model}
    \vspace{-3mm} 
\end{figure*}

Here we consider a MIMO system with $N_t$ transmit and $N_r$ receive antennas with the system model illustrated in Fig~\ref{fig:mimo_model} where $h^{r,t}(\tau,\nu)$ corresponds to DD domain impulse response between the receive antenna $\mathrm{r}$ and transmit antenna $\mathrm{t}$. We use the same transmit and receive pulse shaping filters across all of the antennas.  
The extension of the SISO formulation to the MIMO case can be carried out in two ways. 
The first approach—originally proposed for MC-OTFS, considers the matrix form of the SISO I/O relation in \eqref{sisomat} for each receive-transmit antenna pair $(r,t)$ as
\begin{equation}
\mathbf{y}^{r} = \mathbf{H}^{r,t}\mathbf{x}^{t} + \mathbf{n}^{r}
\end{equation}
Subsequently, all per-antenna channel matrices and signal vectors are stacked vertically 
to form the overall MIMO I/O relation.
\begin{equation}
\label{Zakmmimo}
\resizebox{\linewidth}{!}{$
\underbrace{
\begin{pmatrix}
\mathbf{y}^0  \\ 
\mathbf{y}^1 \\
\vdots  \\
\mathbf{y}^{N_r - 1}
\end{pmatrix}
}_{\text{\large$\mathbf{y}^{\mathrm{MIMO}}$}}
=
\underbrace{
\begin{pmatrix}
\mathbf{H}^{0,0} & \mathbf{H}^{0,1} & \cdots & \mathbf{H}^{0,N_t - 1} \\ 
\mathbf{H}^{1,0} & \mathbf{H}^{1,1} & \cdots & \mathbf{H}^{1,N_t-1} \\
\vdots & \vdots & \vdots & \vdots \\
\mathbf{H}^{N_r-1,0} & \mathbf{H}^{N_r-1,1} & \cdots & \mathbf{H}^{N_r-1,N_t-1} 
\end{pmatrix}
}_{\text{\large$\mathbf{H}^{\mathrm{MIMO}}$}}
\underbrace{
\begin{pmatrix}
\mathbf{x}^0  \\ 
\mathbf{x}^1 \\
\vdots  \\
\mathbf{x}^{N_t-1}
\end{pmatrix}
}_{\text{\large$\mathbf{x}^{\mathrm{MIMO}}$}}
\\
+ 
\underbrace{
\begin{pmatrix}
\mathbf{n}^0  \\ 
\mathbf{n}^1 \\
\vdots  \\
\mathbf{n}^{N_r-1}
\end{pmatrix}
}_{\text{\large$\mathbf{n}^{\mathrm{MIMO}}$}}
$},
\end{equation}
where  $\mathbf{y}^{\rm MIMO}\in\mathbb{C}^{N_rMN\times 1}$, $\mathbf{H}^{\rm MIMO} \in \mathbb{C}^{N_rMN\times N_tMN}$, $\mathbf{x}^{\rm MIMO} \in \mathbb{C}^{N_tMN\times 1}$ and $\mathbf{n}^{\rm MIMO}\in \mathbb{C}^{N_rMN\times 1}$ .
Another perspective can be obtained by revisiting~\eqref{eq:Zak_io}. Each index pair $[k,l]$ in $h_{\mathrm{eff}}$ represents reflections in the propagation environment for which the propagation delay is approximately $k/B$ and
the Doppler shift is approximately $l/T$. When this framework is extended to the MIMO case, the underlying principle remains the same; however, instead of a single transmit–receive antenna pair, multiple antenna pairs are simultaneously coupled through these reflectors. In the SISO setting, the $(k,l)$-th index pair corresponds to the channel coefficient $h_{\rm eff}[k,l]$ which is a scalar complex value, whereas in the MIMO case, each index pair corresponds to a complex $N_r \times N_t$ matrix. Consequently, for each index pair $(k,l)$, we receive a vector
${\bf y}_{\rm dd}[k,l] \in \mathbb{C}^{N_r \times 1}$ which is the vector of received DD samples at the $(k,l)$-th DD bin on all the $N_r$ receive antennas. Extending~\eqref{eq:Zak_io} for the MIMO case we therefore get:
\begin{equation}
\begin{split}
\mathbf{y}_{dd}[k,l]
&= \sum_{k',l' \in \mathbb{Z}}
\mathbf{H}_{\mathrm{eff}}[k',l']
\mathbf{x}_{dd}[k-k',l-l']
e^{j2\pi\frac{l'(k-k')}{MN}} \\
&\quad + \mathbf{n}_{dd}[k,l].
\end{split}
\end{equation}
where $\mathbf{y}_{dd}[k,l] \in \mathbb{C}^{N_r \times 1},\;  \mathbf{H}_{\mathrm{eff}}[k,l]\in \mathbb{C}^{N_r \times N_t}$, $\mathbf{n}_{dd}[k,l]\in \mathbb{C}^{N_r \times 1}$ and $\mathbf{x}_{dd}[k,l]\in \mathbb{C}^{N_t \times 1}$ for $k,l \in \mathbb{Z}$. Here, $\mathbf{x}_{\rm dd}[k,l]$ is the vector of DD domain information symbols transmitted on the $(k,l)$-th DD carrier $\phi_{\rm dd}^{k,l}(\tau,\nu)$ from the $N_t$ transmit antennas. 

Considering the ULA geometric model for the $i$th reflector, the transmit steering vector is given by

\begin{subequations}\label{eq:array_vectors}
\begin{align}
\mathbf{a}_T(\theta_i^{\mathrm{AoD}})
&=
\begin{bmatrix}
1 &
\cdots &
e^{-j2\pi \frac{d_T}{\lambda}(N_t-1)\sin(\theta_i^{\mathrm{AoD}})}
\end{bmatrix}^{T}\in\mathbb{C}^{N_t\times 1},\\[4pt]
\mathbf{a}_R(\theta_i^{\mathrm{AoA}})
&=
\begin{bmatrix}
1 &
\cdots &
e^{-j2\pi \frac{d_R}{\lambda}(N_r-1)\sin(\theta_i^{\mathrm{AoA}})}
\end{bmatrix}^{T}\in\mathbb{C}^{N_r\times 1}.
\end{align}
\end{subequations}
 Here, $\theta_i^{\mathrm{AoD}}$ and $\theta_i^{\mathrm{AoA}}$ denote the angle of departure and angle of arrival of the $i$th path, $d_T$ and $d_R$ are the transmit and receive antenna spacings, and $\lambda$ is the carrier wavelength.
For the $i$th reflector, the channel DD spreading function between the $t$th transmit and $r$th receive antenna is given by the $(r,t)$-th element of the
matrix. The $i$th path of the DD domain physical channel matrix can be written as:
\begin{equation}
\label{geoZak}
    \mathbf{H}_{i}(\tau,\nu) = h_{i} \cdot \mathbf{a}_R(\theta_{i}^{\mathrm{AoA}}) \cdot \mathbf{a}_T^H(\theta_{i}^{\mathrm{AoD}}) \delta(\tau-\tau_{i})\delta(\nu - \nu_{i})
 \in \mathbb{C}^{N_r \times N_t}.
\end{equation}
Based on the formulation in~\eqref{geoZak}, the overall DD spreading function for the channel between the $r$th receive and $t$th transmit antenna is given by the $(r,t)$-th
element of the matrix:
\begin{equation}
    \mathbf{H}_{\mathrm{phy}}(\tau,\nu) = \sum_{i = 1}^{L}\mathbf{H}_{i}(\tau,\nu).
\end{equation}
Assuming $h_{\mathrm{phy}}^{r,t}(\tau,\nu) $ is the element of the $\mathrm{r}$th row and $\mathrm{t}$th column of $\bf H_{\mathrm{phy}}(\tau,\nu)$, the effective continouous DD domain channel filter between the $t$th transmit and $r$th receive antenna is given by:
\begin{equation}
    h_{\mathrm{eff}}^{r,t}(\tau,\nu) = 
w_{\mathrm{rx}}(\tau,\nu) 
\ast_{\sigma} h_{\mathrm{phy}}^{r,t}(\tau,\nu) 
\ast_{\sigma} w_{\mathrm{tx}}(\tau,\nu)
\end{equation}

\begin{figure}[!h]
    \centering

    \includegraphics[
        trim = 0cm 3cm 0cm 3cm,
        clip,
        width=1\columnwidth,
        height=0.72\columnwidth,
        keepaspectratio
    ]{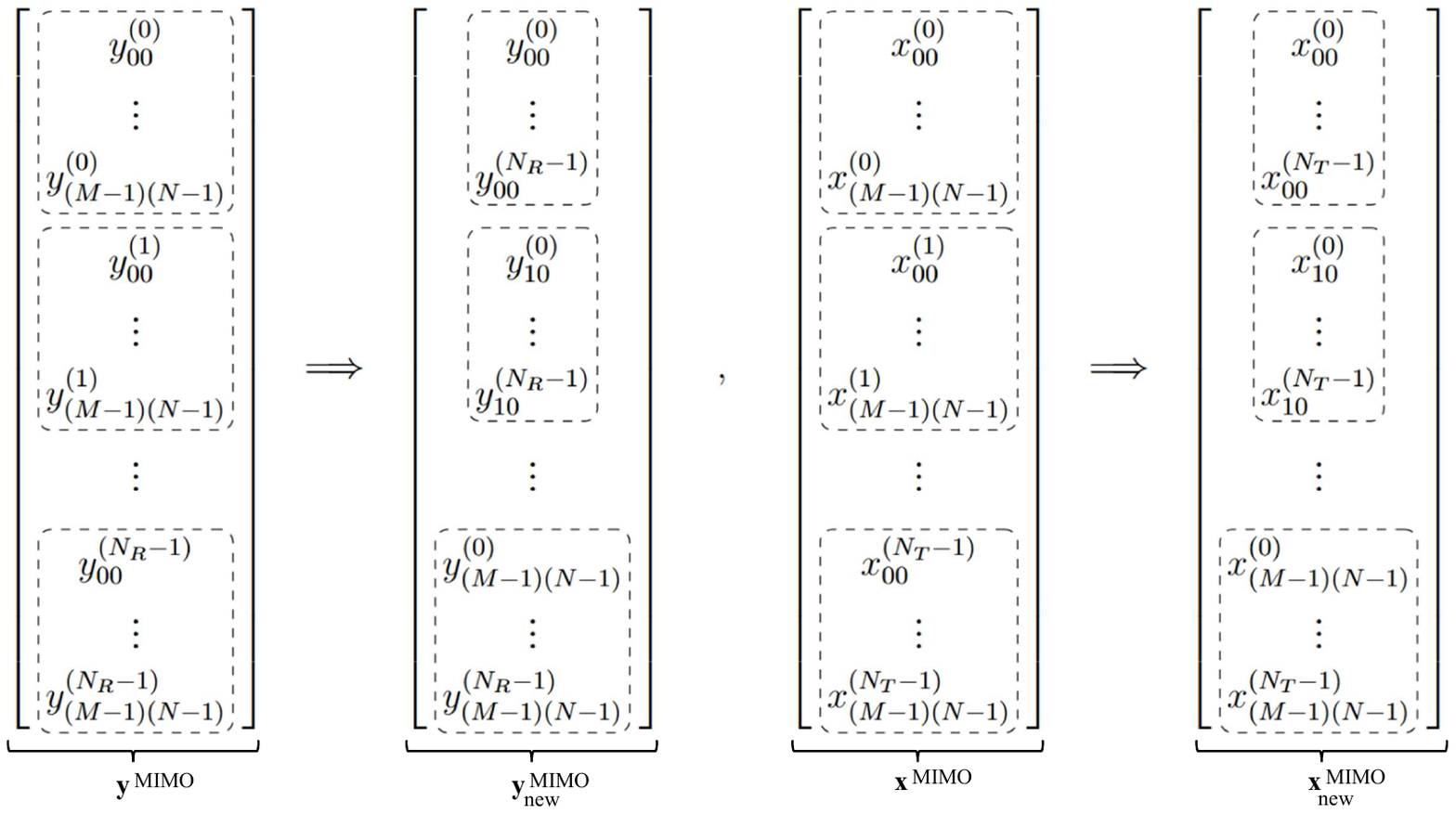}

    \vspace{2mm}
    \includegraphics[
        trim = 4cm 8cm 3cm 7.5cm,
        clip,
        width=1\columnwidth,
        height=1\columnwidth,
        keepaspectratio
    ]{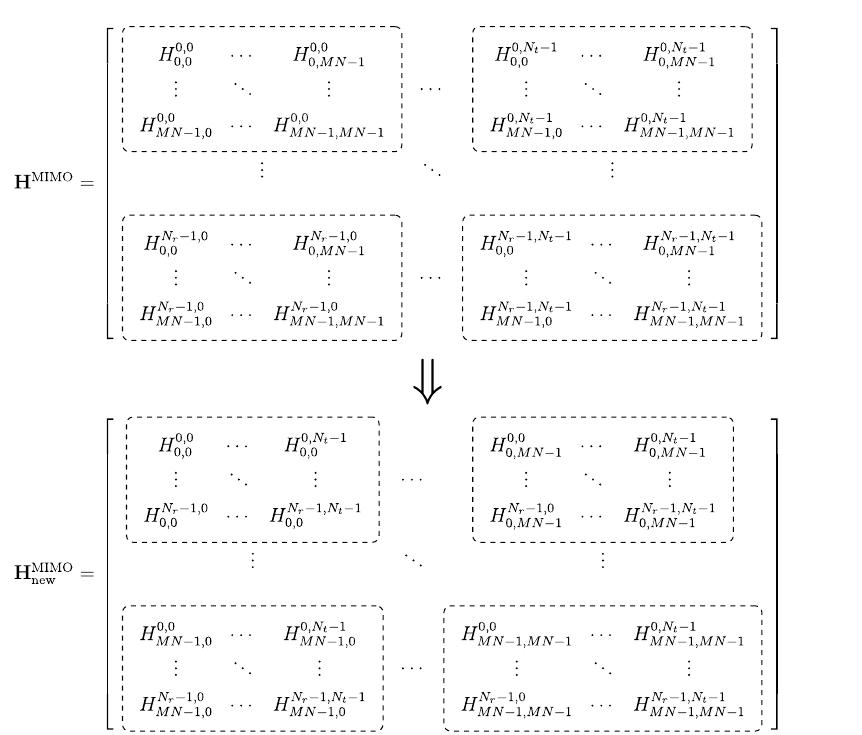}

    \caption{Permutation method.}
    \label{fig:perm}

    \vspace{-3mm}
\end{figure}

The received match-filtered continuous DD domain signal at the $r$th receive antenna is given by
$y_{r\rm,dd}^{w_{\rm rx}}(\tau, \nu) = \sum\limits_{t=0}^{N_t-1} h_{\rm eff}^{r,t}(\tau, \nu) \ast_{\sigma} x_{t,\rm dd}^{w_{\rm tx}}(\tau, \nu) \, + \, n_{r,\rm dd}^{w_{\rm rx}}(\tau, \nu)$
where $n_{r,\rm dd}^{w_{\rm rx}}(\tau, \nu) = w_{\rm rx}(\tau, \nu) \ast_{\sigma} n_{r\rm, dd}(\tau, \nu)$ and $n_{r,\rm dd}(\tau, \nu)$ is the DD representation (Zak transform) of the time-domain AWGN at the $r$th receive antenna. Sampling $y_{r\rm ,dd}^{w_{\rm rx}}(\tau, \nu)$ on the information lattice gives $y_{r,\rm dd}[k,l]$. Clearly the I/O relation between $y_{r,\rm dd}[k,l]$ and the transmitted DD symbols $x_{t,\rm dd}[k,l]$, $t=0,1,\cdots, N_t-1$ is given by $y_{r,\rm dd}[k,l] = \sum\limits_{t=0}^{N_t-1} \sum\limits_{k', l' \in \mathbb{Z}} h_{\rm eff}^{r,t}[k',l'] \, x_{t,\rm dd}[k - k', l - l'] \, e^{j 2 \pi l' (k - k')/MN} \,
 \, + n_{r,\rm dd}[k,l] $ where $h_{\rm eff}^{r,t}[k,l]$ is simply $h_{\rm eff}^{r,t}(\tau, \nu)$ sampled on the information lattice, and $n_{r,\rm dd}[k,l]$ is simply $n_{r,\rm dd}^{w_{\rm rx}}(\tau,\nu)$ sampled on the information lattice.
We organize the $MN$ received DD samples at the $r$th receive antenna $y_{r,\rm dd}[k,l]$, $k=0,1,\cdots, M-1, l=0,1,\cdots, N-1$ into the vector ${\bf y}^{(r)} \in C^{MN \times 1}$. Stacking the received vectors for all $N_r$ antennas we get the vector $\mathbf{y}_{\mathrm{new}}^{\mathrm{MIMO}}
=
\left[
(\mathbf{y}^{(0)})^{T},
(\mathbf{y}^{(1)})^{T},
\ldots,
(\mathbf{y}^{(N_r-1)})^{T}
\right]^{T}
\in
\mathbb{C}^{N_rMN \times 1}
$. Clearly, this received vector is simply a reshuffling of the elements of the vector $\mathbf{y}^{\rm MIMO}$ in~\eqref{Zakmmimo} (see also Fig~\ref{fig:perm}).
Similarly, we also organize the $MN$ DD symbols $x_{t,\rm dd}[k,l]$ on all $N_t$ transmit antennas into a vector ${\bf x}_{\rm new}^{\rm MIMO} \in \mathbb{C}^{N_tMN \times 1}$ which is also a permutation of the elements of $\mathbf{x}^{\rm MIMO}$ in~\eqref{Zakmmimo}.
We therefore have $\bf y_{\rm new}^{\rm MIMO} = H_{\rm new}^{\rm MIMO} x_{\rm new}^{\rm MIMO} + \mathbf{n}_{\rm new}^{\rm MIMO}$, where:
\begin{subequations}
\begin{align}
\mathbf{y}_{\mathrm{new}}^{\mathrm{MIMO}} &= \mathbf{P}_y \, \mathbf{y}^{\mathrm{MIMO}}, \label{eq:permY} \\[2mm]
    \mathbf{n}_{\mathrm{new}}^{\mathrm{MIMO}} &= \mathbf{P}_y \, \mathbf{n}^{\mathrm{MIMO}}, \label{eq:permn} \\[2mm]
    \mathbf{x}_{\mathrm{new}}^{\mathrm{MIMO}} &= \mathbf{P}_x \, \mathbf{x}^{\mathrm{MIMO}}, \label{eq:permX} \\[2mm]
    \mathbf{H}_{\mathrm{new}}^{\mathrm{MIMO}} &= \mathbf{P}_y \, 
    \mathbf{H}^{\mathrm{MIMO}} \, \mathbf{P}_x^{\mathsf{T}}. \label{eq:permH}
\end{align}
\end{subequations}

Here $\mathbf{P}_y$ and $\mathbf{P}_x$ are appropriate permutation matrices.

\section{Methodology}
\label{sec:3}
\begin{figure*}[!t]
\centering
\scalebox{0.9}{
\begin{minipage}{\textwidth}
\centering

\subfloat[SISO\label{fig:4a}]{\vspace{-14mm}
  \includegraphics[trim = 6cm 4cm 5.5cm 5cm,clip,width=0.48\textwidth]{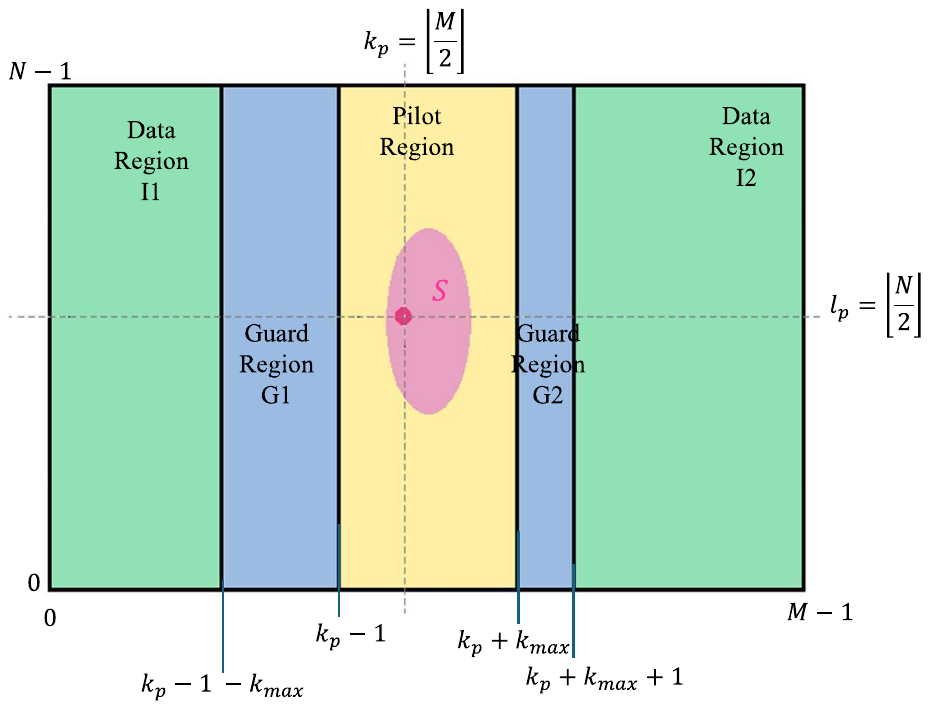}
}%
\hfill
\subfloat[Separation in Doppler only\label{fig:4b}]{\vspace{-14mm}
  \includegraphics[trim = 6cm 4cm 5cm 5cm,clip,width=0.48\textwidth]{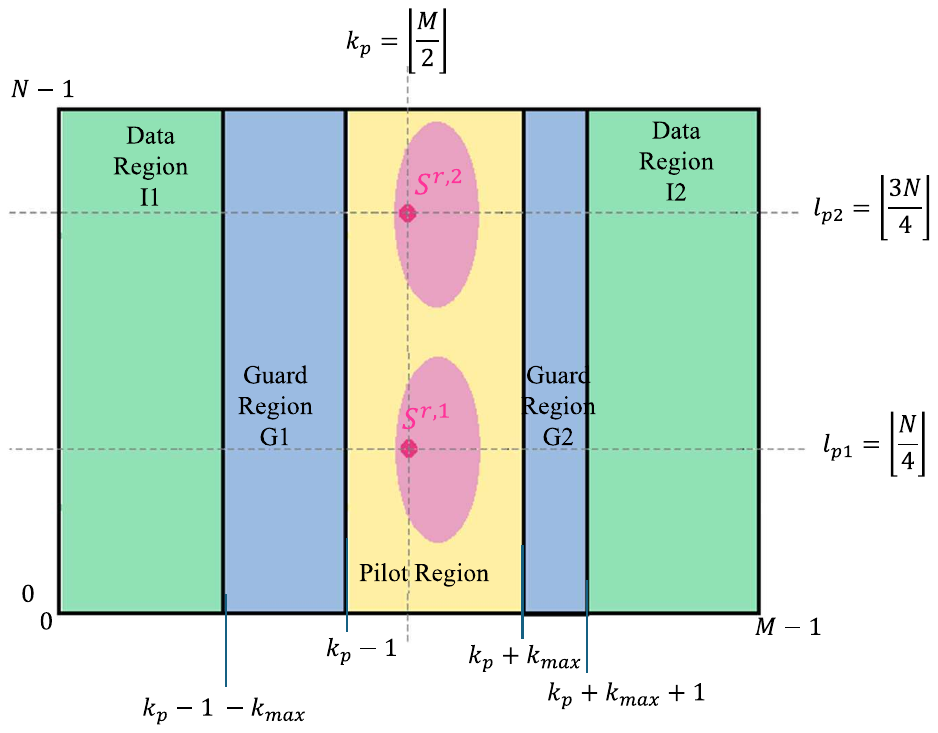}
}\\[2ex]
\vspace{-7mm}
\subfloat[Separation in Delay only\label{fig:4c}]{
  \includegraphics[trim = 6cm 4cm 5cm 5cm,clip,width=0.48\textwidth]{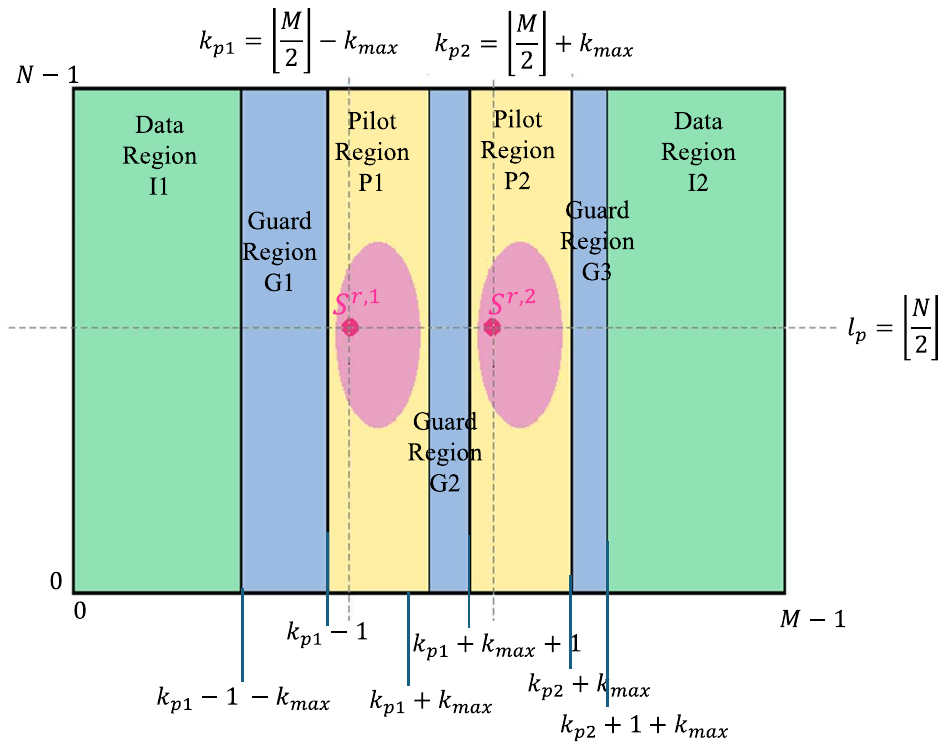}
}
\hfill
\subfloat[Separation in both Delay and Doppler\label{fig:4d}]{
  \includegraphics[trim = 6cm 4cm 5.5cm 4.5cm,clip,width=0.48\textwidth]{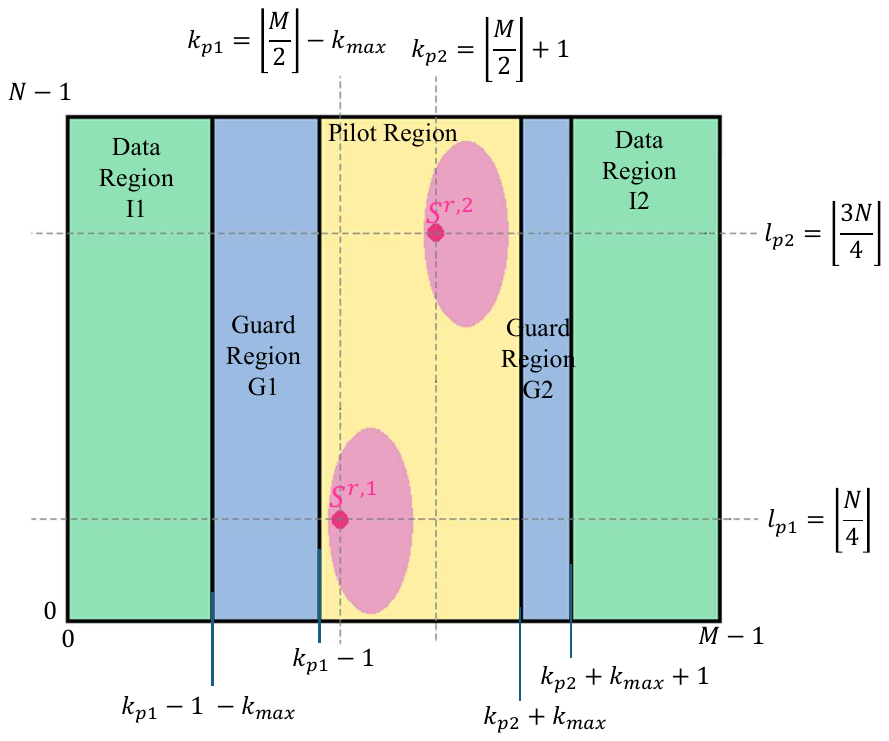}
}\\[2ex]
\end{minipage}
}
\caption{Pilot separation strategies in the delay--Doppler (DD) plane: 
(a) SISO, 
(b) separation in Doppler only, 
(c) separation in delay only, and 
(d) separation in both delay and Doppler. 
The pilot, guard, and data regions are denoted by 
$\mathcal{P}=\bigcup_i P_i$, 
$\mathcal{G}=\bigcup_i G_i$, and 
$\mathcal{I}=\bigcup_i I_i$, respectively.}
\label{fig:fig4}
\end{figure*}

In SISO Zak-OTFS, the channel filter $h_{\rm eff}[k,l]$ is estimated from the channel response to a single DD pulse (pilot) transmitted at the $(k_p, l_p)$ (i.e., the DD carrier is $\phi^{k_p, l_p}_{\rm dd}(\tau, \nu)$ depicted as a red dot in Fig~\ref{fig:4a}). The channel response is received
on the DD taps within the pink ellipse shown with support set $\mathcal{S}$ in Fig~\ref{fig:4a}. The DD spread of this channel response ellipse is same as the spread of the effective channel filter $h_{\rm eff}[k,l]$ which is simply the channel delay spread along the delay axis (horizontal) and the channel Doppler spread along the vertical Doppler axis. Due to quasi-periodicity, the pilot pulse repeats with period $M$ and $N$ along the delay and Doppler axis respectively. The channel response to the periodic repetitions do not overlap/alias if the channel delay and Doppler spread are less than the respective periods. Under this condition, known as the crystallization condition \cite{mcvsZak}, the taps of $h_{\rm eff}[k,l]$ can be estimated simply from the response received within the pink ellipse shown in Fig~\ref{fig:4a}~\cite{mcvsZak}.
To avoid interference between the data DD pulses and the pilot DD pulse, a guard region is provided between the pilot region and the data region (data region consists of DD pulses used for data transmission). The guard region (blue in Fig~\ref{fig:4a}) does not carry any data pulses and is therefore an overhead.

For the MIMO extension, since the crystallization condition is determined by the underlying physical channel, its satisfaction in the SISO case implies it also holds for the MIMO configuration. The challenge is therefore the design of pilot transmission for channel estimation. In particular, the channel response to the pilots transmitted from different transmit antennas must not interfere.
To achieve interference free reception of pilot response from different transmit antennas, we may need to locate the DD pulses corresponding to the pilots transmitted from different antennas sufficiently far apart in the DD domain. This results in a higher pilot overhead compared to SISO Zak-OTFS. There is therefore
a trade-off between pilot overhead and interference between channel response to pilots transmitted from different transmit antennas. To achieve this trade-off, one can consider at least three pilot-separation strategies in the DD domain illustrated in Fig~\ref{fig:fig4}.

We achieve interference free pilot response through separation of the DD location of the pilots transmitted from different transmit antennas. This can be separation in  Doppler (Fig~\ref{fig:4b}), delay (Fig~\ref{fig:4c}), or jointly in both (Fig~\ref{fig:4d}). 
When separation is applied in the delay only the pilot regions incur approximately twice the overhead; however, the guard regions need not be doubled, as the guard region between adjacent pilot regions can be shared and need not be doubled. 
Alternatively, interference-free pilot response can be achieved by separation only in the Doppler domain. 
In this case, the overall pilot overhead remains unchanged compared to the SISO case, but this approach introduces certain limitations. 
It performs well when highly localized pulse-shaping filters such as Gaussian are used, however, for less localized pulse-shaping filters such as Sinc or root-raised-cosine (RRC), the performance may degrade. 
Moreover, compared to the other two separation methods, one can speculate that the performance of Doppler only separation will degrade for smaller values of Doppler spread and might not scale effectively with an increasing number of transmit antennas. 
Finally, separation in both delay and Doppler provides a balanced design. With only a modest increase in the pilot overhead compared to SISO-Zak-OTFS, it supports a larger number of transmit antennas while preserving interference-free pilot responses.
\vspace{-2mm}
\vspace{2mm}
Each antenna transmits its own data symbols over all designated data regions, 
while pilot symbols are placed only at the antenna's assigned pilot locations. 
No transmission occurs in the guard regions.

For transmit antenna $t$, the data and pilot pulsones are then combined as follows:
\begin{equation}
    x_{t,\mathrm{dd}}[k,l]
= 
\underbrace{
\sqrt{E_{d}^t}\, x_{t,d,\mathrm{dd}}[k,l]
}_{t\text{'th antenna data pulsones}}
+
\underbrace{
\sqrt{E_{p}^t}\, x_{t,p,\mathrm{dd}}[k,l]
}_{t\text{'th antenna pilot pulsone}},
\label{eq:xdd_composition}
\end{equation}
where $E^t_d$ and $E^t_p$ are defined as the average data energy and average pilot energy per transmit antenna respectively, i.e.:
\begin{subequations}\label{eq:EdtEpt}
\begin{align}
\mathbb{E} 
\!\left[
\sum_{k=0}^{M-1}
\sum_{l=0}^{N-1}
\left|
x_{t,d,\mathrm{dd}}[k,l]
\right|^{2}
\right]
&= E^t_d,
\label{eq:Edt}\\[2mm]
\mathbb{E} 
\!\left[
\sum_{k=0}^{M-1}
\sum_{l=0}^{N-1}
\left|
x_{t,p,\mathrm{dd}}[k,l]
\right|^{2}
\right]
&= E^t_p,
 \label{eq:Ept}
\end{align}
\end{subequations}
The overall data and pilot energies can be obtained by summing the per-antenna energies as
\begin{subequations}\label{eq:EdEp}
\begin{align}
E_d &= \sum_{t=0}^{N_t-1} E_d^{(t)}, &
E_p &= \sum_{t=0}^{N_t-1} E_p^{(t)}. 
\end{align}
\end{subequations}

The noise energy at each receive antenna can be defined as:
\begin{equation}
\mathbb{E}\left[\sum_{k=0}^{M-1}\sum_{l=0}^{N-1}|n_{r,\mathrm{dd}}[k,l]|^{2}\right]= M N N_{0}.
\label{eq:noise}
\end{equation}
For the receive transmit antenna pair $(r,t)$, the support set $\mathcal{S}^{r,t}$ of $h^{r,t}_{\mathrm{eff}}[k,l]$ consists of all pairs $(k,l)$ for which $h^{r,t}_{\mathrm{eff}}[k,l] \neq 0$. Then the effective channel energy gain can be expressed as $\displaystyle \sum_{(k,l)\in \mathcal{S}^{r,t}} \!\! \left| h^{r,t}_{\mathrm{eff}}[k,l] \right|^{2}$,
and therefore the average received pilot and data energy at the $r$th receive antenna from the $t$th transmit antenna are: 
$E^t_{d} \! \sum_{(k,l)\in \mathcal{S}^{r,t}} \! \left| h^{r,t}_{\mathrm{eff}}[k,l] \right|^{2}$
and $E^t_{p} \! \sum_{(k,l)\in \mathcal{S}^{r,t}} \! \left| h^{r,t}_{\mathrm{eff}}[k,l] \right|^{2}$,
respectively. 

Hence, the ratio of the power of the received data pulsones to the noise power
(i.e., the data SNR) at the $r$'th receive antenna is given by
\begin{equation}
\rho^r_{d}
\triangleq
\sum_{t = 0}^{N_t-1}\frac{
E^t_{d}
\displaystyle\sum_{(k,l)\in \mathcal{S}}
\left| h^{r,t}_{\mathrm{eff}}[k,l] \right|^{2}
}{
M N N_{0}
}.
\label{eq:data_snr}
\end{equation}

Similarly, the ratio of the power of the received pilot pulsone to that of noise
(i.e., the pilot SNR) is given by
\begin{equation}
\rho^r_{p}
\triangleq
\sum_{t = 0}^{N_t-1}\frac{
E^t_{p}
\displaystyle\sum_{(k,l)\in \mathcal{S}}
\left| h^{r,t}_{\mathrm{eff}}[k,l] \right|^{2}
}{
M N N_{0}
}.
\label{eq:pilot_snr}
\end{equation}
Since the effective channels corresponding to different transmit–receive antenna pairs differ only in phase arising out of the ULA geometry, the resulting SNR values are identical across all receive antennas. In other words, the overall data and pilot SNRs are equal to their per-antenna counterparts, i.e.,
\(\rho_{d} = \rho_{d}^{r}\) and \(\rho_{p} = \rho_{p}^{r}\).
Hence, the ratio of the received pilot power to the received data power is given by
\(\rho_{p}/\rho_{d}\),
which is referred to as the \emph{pilot-to-data power ratio (PDR)}.
To ensure a fair comparison and consistent total transmit power across antennas, 
the transmit power is normalized at the transmitter side. Specifically, for each transmit antenna, the pilot energy is set to $E^t_{p} = E_p/N_{t}$, while the data energy $E^t_{d}  =E_d/N_t$ is determined according to the chosen value of the PDR.\\
With this normalization, the effective discrete DD-domain channel filter $h^{r,t}_{\mathrm{eff}}[k,l]$ 
for each transmit--receive antenna pair is estimated based on the received pilot symbols 
within the pilot region $\mathcal{P}$. 
The maximum-likelihood (ML) estimator is obtained from the samples of the
cross-ambiguity function between the received pilot pulsone \( y_{r,p,\mathrm{dd}}[k,l] \)
and the transmitted pilot pulsone \( x_{t,p,\mathrm{dd}}[k,l] \), when observed only within
the support set \(\mathcal{S}^{r,t}\)~\cite{mcvsZak}.
For \((k,l) \in \mathcal{S}^{r,t}\), we have
\begin{align}
\hspace{1.2cm}\hat{h}^{r,t}_{\mathrm{eff}}[k,l]
&= A_{y_{r,p},x_{t,p}}[k,l] \notag\\[-2pt]
&\hspace{-2cm}= \sum_{k'=0}^{M-1}\sum_{l'=0}^{N-1}
y_{r,p,\mathrm{dd}}[k',l']\,x_{t,p,\mathrm{dd}}^{*}[k'-k,\,l'-l]\,
e^{-j2\pi \frac{l (k'-k)}{MN}}.
\label{eq:ml_estimator}
\end{align}

The resulting estimate is then used to construct an estimate of the effective DD domain channel matrix $\bf H^{\rm MIMO}$ in~\eqref{Zakmmimo}. For detecting the transmitted information symbols,
we use the matrix-vector I/O relation in~\eqref{Zakmmimo} with only those elements of $\bf y^{\rm MIMO}$ which correspond to DD locations outside the pilot regions. Similarly, for $\bf x^{\rm MIMO}$ we only consider those elements which correspond to DD locations in the data regions $\mathcal{I}$. The effective channel matrix is then a sub-matrix of $\bf H^{\rm MIMO}$ with only those rows which correspond to DD locations outside the pilot regions and only those columns which correspond to DD locations in the data region.
We use an MMSE equalizer to equalize the I/O relation in~\eqref{Zakmmimo}.\\
Joint equalization of all information symbols transmitted across multiple antennas has higher complexity compared to per-carrier MIMO equalization in CP-OFDM, although recently for SISO Zak-OTFS low complexity equalization methods
have been proposed with complexity linear in the number of information symbols (similar to complexity of per-carrier equalization)~\cite{comp2}.

\section{Simulations and Results}
For simulation purposes, we consider the CDL-C channel model, the power and delay profile of which has been specified in the Table 7.7.1-3 described in~\cite{CDL-B}. 
The Doppler shift of each multi-path component is modeled geometrically based on the relative motion between the user equipment (UE) receiver and the direction of wave propagation. The UE velocity direction in the horizontal plane is modeled as a unit vector $
\mathbf{u}_{\text{UE}} =
\begin{bmatrix}
\cos \phi &\sin \phi & 0
\end{bmatrix}^T,
$
where $\phi \sim [0,2\pi]$ is uniformly distributed.
For the $p$-th propagation path with angle of arrival $\mathrm{AoA}_p$, the wave propagation direction is represented by
$
\mathbf{u}_{p} =
\begin{bmatrix}
\cos(\mathrm{AoA}_p) &
\sin(\mathrm{AoA}_p) & 0
\end{bmatrix}^T.
$

The Doppler shift of the $p$-th path is then computed as the projection of the UE receiver velocity vector onto the wave propagation direction,
\begin{equation*}
f_{D,p} = \nu_{\max} \, \langle\mathbf{u}_{p}, \mathbf{u}_{\text{UE}}\rangle .   
\end{equation*}
This model captures the fact that each multi-path component experiences a different Doppler shift depending on its angle of arrival relative to the UE motion. This geometry-based Doppler modeling ensures consistency between the spatial characteristics of the channel and its temporal variation.
We consider a SISO and a $2 \times 2$ MIMO Zak-OTFS system. For comparison, a CP-OFDM system with the same bandwidth and frame duration is also considered. 
The system parameters have been specified in Table~\ref{tab:param}.
For standardized 3GPP 5G NR CP-OFDM, we maintain a fixed time-bandwidth product while scaling the OFDM symbol duration and the number of slots according to the SCS. For normal CP configurations (i.e., SCS $= 15$ and $30$ kHz), each slot contains 14 OFDM symbols. For extended CP (used for SCS $= 60$ kHz), each slot contains 12 OFDM symbols.

We optimize the effective spectral efficiency (SE) over all possible Type-A pilot configurations defined in 3GPP 5G NR for all Modulation and Coding Schemes (MCS) specified in MCS Table index~1 (Table 5.1.3.1-1 in~\cite{ref28}). The LDPC codeword spans all OFDM symbols within the considered CP-OFDM subframe. For a given subcarrier spacing, pilot configuration, and MCS, the effective SE is defined as
$\mathrm{SE}_{\mathrm{eff}} = (1 - \mathrm{BLER}) \frac{N_{\mathrm{info}}}{B \cdot T_{\rm frame}},$
where $N_{\mathrm{info}}$ denotes the number of information bits and $T_{\rm frame }$ refers to the total effective frame time. In Zak-OTFS, the effective frame duration is given by $T_{\rm frame}=T+\tau_{\max}$, where the additional $\tau_{\max}$ term accounts for the guard interval between consecutive frames. In contrast, for CP-OFDM, the cyclic prefix already accommodates the maximum channel delay spread, and therefore the effective frame duration is simply $T_{\rm frame}=T$. Here, BLER denotes the average block error rate. 

Both systems employ MMSE equalization for a fair comparison, with OFDM being per-subcarrier spatial equalization. The pulse-shaping filter is chosen between a Sinc filter $\mathrm{w}_{\mathrm{tx}}(\tau,\nu) = \sqrt{BT}\mathrm{sinc}(B\tau)\mathrm{sinc}(T\nu)$ or a Gauss-Sinc filter $ \mathrm{w}_{\mathrm{tx}}(\tau,\nu) = \Omega_{\tau}\Omega_{\nu}\sqrt{BT}\mathrm{sinc}(B\tau)\mathrm{sinc}(T\nu)e^{-\alpha_{\tau}B^2\tau^2}e^{-\alpha_{\nu}T^2\nu^2}$ with parameters $\Omega_{\tau} = \Omega_{\nu} = 1.0278$, $\alpha_{\tau} = \alpha_{\nu} = 0.044$ to ensure that $99\%$ of the energy is captured in bandwidth B and time duration T.
\begin{table}[h]
    \centering
    \caption{Simulation Parameters}
    \begin{tabular}{|c|c|}
    \hline
        Carrier Frequency & $f_c = 7~\rm GHz$\\ \hline
        Frame Duration & $T = 1~\rm ms$ \\ \hline
        Delay Profile & Normal with $\tau_{\rm DS} = 363 \, ns $\\ \hline
         Bandwidth & $B = 720~\rm kHz$ \\ \hline
         Maximum Speed & $v_{\rm max} = 350~\rm Km/h$ \\ \hline
         Doppler Period & $\nu_p = 10~\rm kHz$ \\ \hline
         $(M,N)$ & $(72,10)$ \\ \hline
        Half Wavelength Antenna Spacing & $d_T = d_R = \frac{\lambda}{2}$ \\ \hline
         Coding & LDPC \\  \hline    
         BLER Threshold & $0.1$ \\ \hline
         Supported Subcarrier Spacing & SCS $ \in \{15, 30, 60\}$ kHz \\ \hline
        Pulse Shaping Filter & Sinc and Gauss-Sinc \\ \hline
        Cyclic Prefix & Normal and Extended\\ \hline
        Normal CP length over all slots & $4.7$ and $5.138~\rm \mu s$
        \\ \hline
        OFDM Pilot Configuration & Type-A/comb structure (see 7.4.1.7.3 in~\cite{3gpp38211}) \\ \hline

   \end{tabular}
    
    \label{tab:param}
\end{table}

\subsection{BER Under Perfect CSI}
\begin{figure}[h]
    \centering
    \includegraphics[width=\columnwidth]{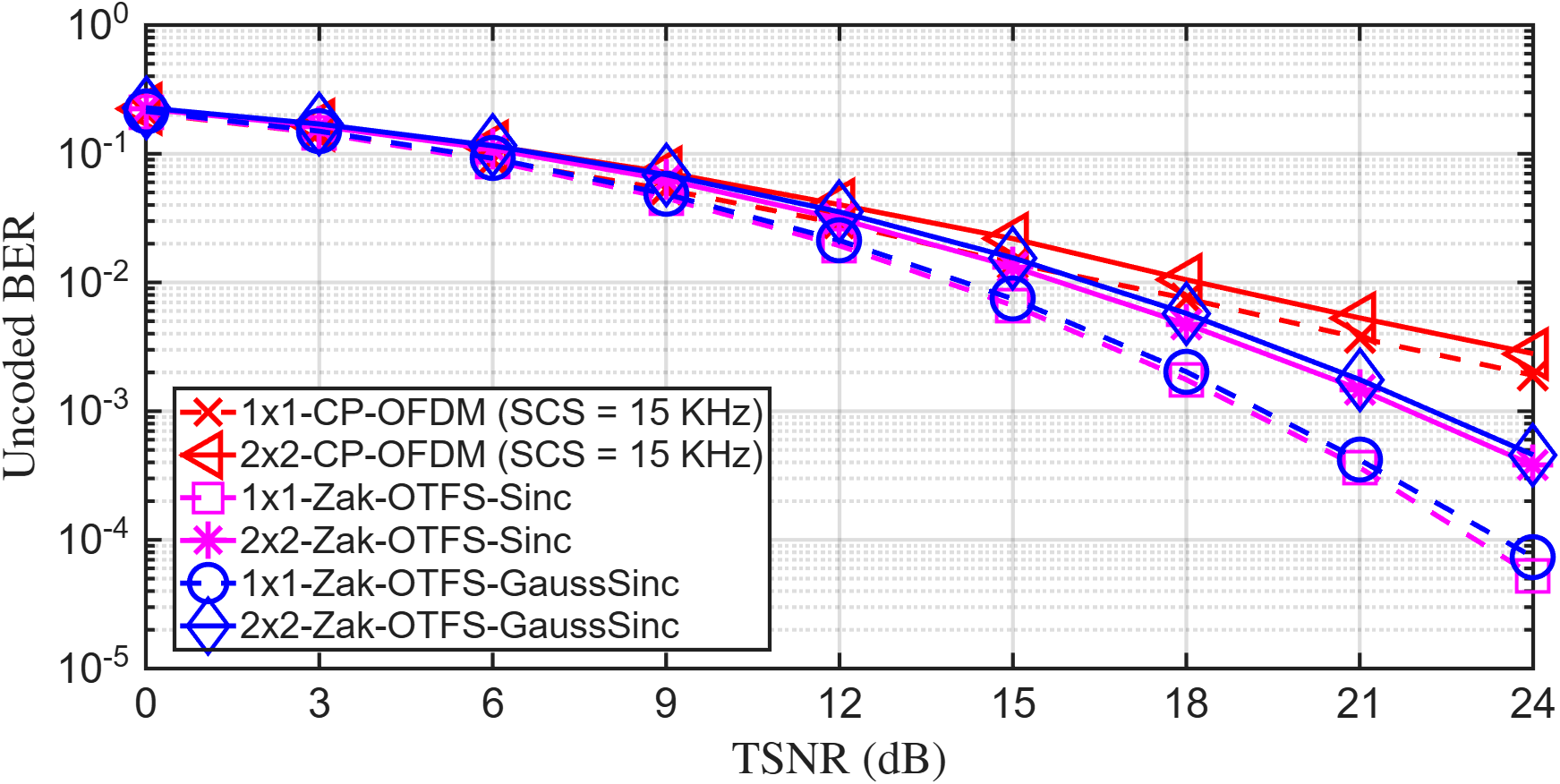}
    \caption{Uncoded 4QAM BER vs TSNR for Perfect CSI with $\nu_{\rm max} =600~\rm Hz$.}
    \label{fig:perfect_CSI}
\end{figure}
One of the main advantages of Zak-OTFS over OFDM, as observed in the SISO case, is its ability to mitigate channel fading, leading to improved performance~\cite{otfs_book}. These gains are expected to naturally extend to MIMO systems. Fig~\ref{fig:perfect_CSI} illustrates the BER vs Total SNR (TSNR) performance under the assumption of perfect CSI for a maximum Doppler shift $\nu_{max} = 600~\mathrm{Hz}$ and 4QAM modulation. TSNR is the ratio of the total power of data and pilot carriers to the noise power and is equal to $\left( \rho_p + \rho_d \right)$. For benchmarking this specific case with OFDM we consider only the $15~\mathrm{kHz}$ subcarrier spacing, as all numerologies exhibit similar performance under the perfect CSI assumption and $\nu_{max} = 600$ Hz.
In this scenario, we observe that both SISO and MIMO Zak-OTFS outperform their corresponding CP-OFDM counterparts. This indicates that CP-OFDM is more susceptible to channel fading in time and frequency, while Zak-OTFS maintains greater robustness under such conditions. 
\subsection{MIMO Zak-OTFS Pilot Configuration and Channel Estimation}
Moving beyond the assumption of perfect CSI, an appropriate pilot placement strategy must be designed for reliable channel estimation. As mentioned in Section~\ref{sec:3}, the objectives are twofold: first, to ensure sufficient isolation of response to pilots transmitted from different transmit antennas for accurate channel estimation; and second, to achieve this isolation without incurring a linear increase in pilot overhead with increasing number of transmit antennas or inefficient use of resources. We show that slightly separating pilots in both delay and Doppler will achieve this objective.\\
\begin{figure}[h]
    \centering
    \includegraphics[width=\columnwidth]{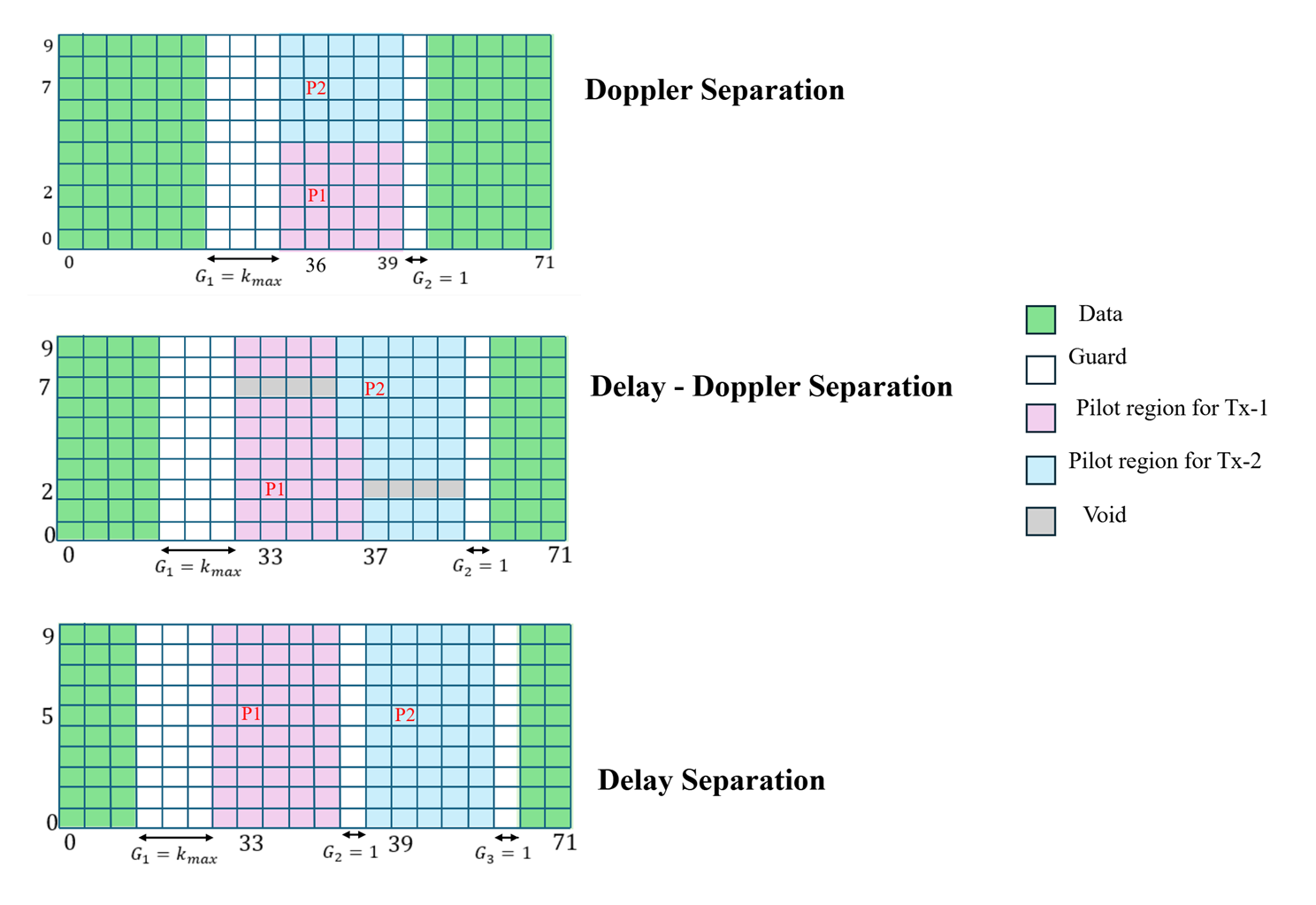}
    \caption{Different Pilot Placement Strategies comprising Doppler, Delay and Delay-Doppler Separations}
    \label{fig:pilot_placement}
\end{figure}
\\
Fig~\ref{fig:pilot_placement} illustrates a more precise pilot and guard-region allocation strategy for the received signal at each receiver. The vertical axis represents the Doppler index, while the horizontal axis denotes the delay index. In the separation method applied jointly in both delay and Doppler, rather than truncating the pilot region only along the Doppler dimension as in the Doppler-only separation scheme, void regions are introduced such that the entire strip is utilized, excluding the single row occupied by the other pilot. This design is motivated by the fact that, although Gauss--sinc filters exhibit strong localization, they still retain significant energy leakage along the row and column corresponding to the DD location of the transmitted pilot. 

\begin{figure}[h]
    \centering
    \includegraphics[width=\columnwidth]{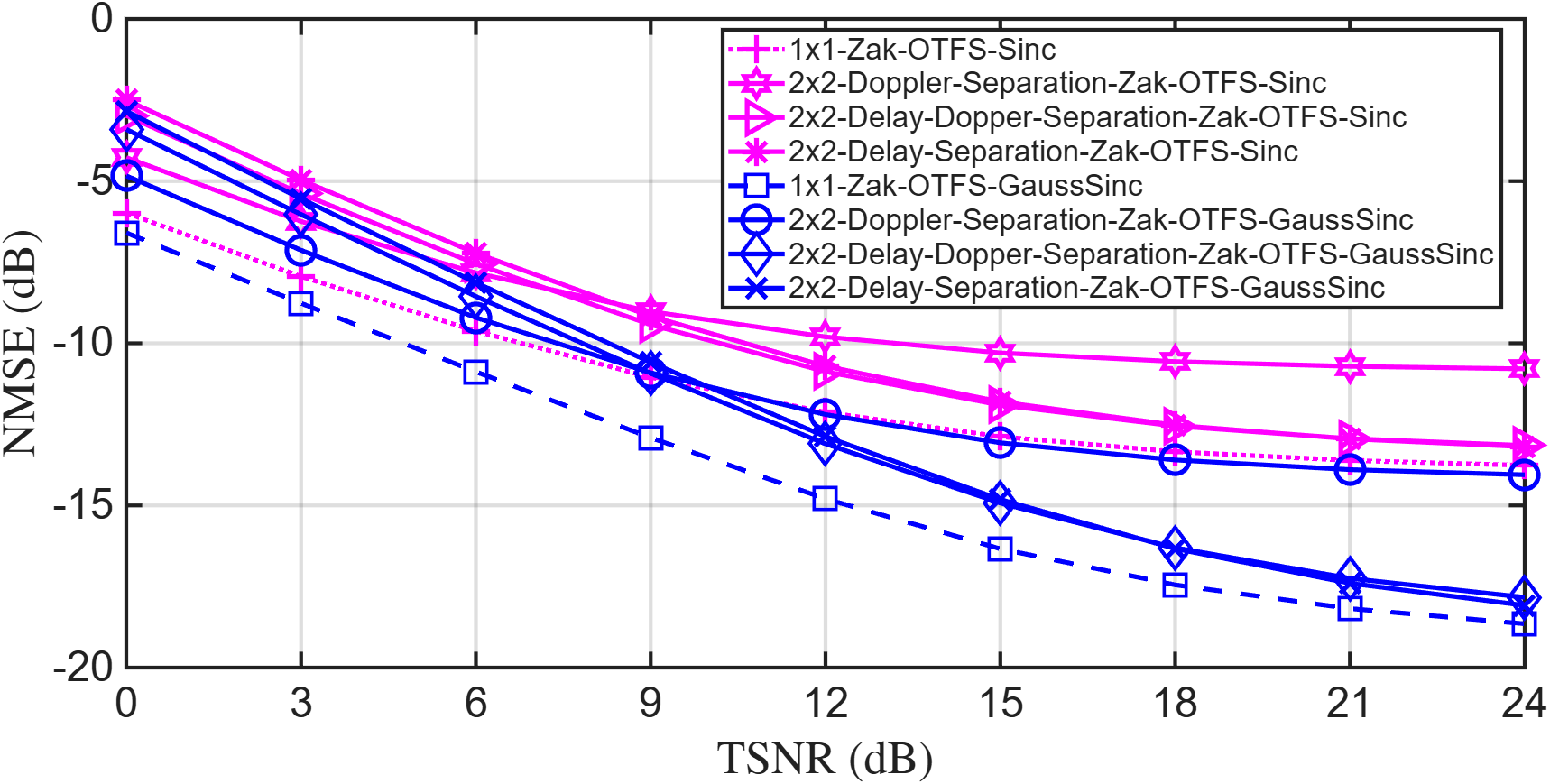}
    \caption{NMSE vs TSNR for Different Pilot Placement Strategies with $\nu_{\rm max} = 1200~\rm Hz$, PDR $=-5~\rm dB$.}
    \label{fig:nmse}
\end{figure}

Since the focus is on higher Doppler scenarios, Fig~\ref{fig:nmse} evaluates the NMSE performance of the two considered pulse-shaping filters at $\nu_{\max}=1200~\mathrm{Hz}$ where:
\begin{equation*}
   \text{NMSE}= \mathbb{E}\left\{\frac{\|\mathbf{H}^{\rm MIMO} - \mathbf{\hat{H}}^{\rm MIMO}\|_F^2}{\|\mathbf{H}^{\rm MIMO} \|_F^2}\right\},
\end{equation*}
where $\hat{\mathbf{H}}^{\rm MIMO}$ is the estimated MIMO channel.

The results first show that the Gauss--sinc filter provides a significantly more accurate channel estimate than the conventional sinc filter. The BER degradation associated with the latter will be demonstrated in the following results.

More importantly, the estimates obtained under different pilot placement strategies can be compared. Assigning two separate strips to different pilots (i.e., Delay separation) offers only marginal improvement relative to a slightly staggered placement within a somewhat wider shared strip (i.e., Delay-Doppler separation). In contrast, separation only along the Doppler dimension leads to noticeable degradation at high Doppler shifts. This is due to interference between the channel response to the pilots transmitted from different transmit antennas.

The NMSE of the $2 \times 2$ MIMO with the Delay-Doppler separation pilot placement is close to the NMSE for the SISO channel, which suggests that the interference between the response to the pilots transmitted from the different transmit antennas is small. Another important factor is pilot overhead. Table~\ref{tab:nmse_increase} compares the pilot overhead for the three considered pilot separations. The results indicate that  Delay separation (i.e., allocating separate strips) provides only limited performance gains while incurring additional pilot overhead. Consequently, its benefit in terms of spectral efficiency is limited. However, at low Doppler shifts where channel response of pilots transmitted from different transmit antennas does not interfere, using only  Doppler separation becomes advantageous due to its lower pilot overhead. As long as inter-pilot interference remains negligible, Doppler separation yields the best spectral efficiency performance.

\begin{table}[h!]
\centering
\caption{Pilot Overhead for Different Separation Methods}
\renewcommand{\arraystretch}{1.2}
\begin{tabular}{|c|c|}
\hline
\textbf{Separation Strategy} & \textbf{Pilot Overhead (\%)} \\
\hline
Doppler only & 12.5 \\
Delay-Doppler  & 18.05 \\
Delay only & 20.8333 \\
\hline
\end{tabular}
\label{tab:nmse_increase}
\end{table}
 Based on these findings, the remainder of this work focuses only on the two relevant strategies: separation in Doppler only, and separation in both delay and Doppler.
 \subsection{BER Under Imperfect CSI}
First, we show that the advantages observed under perfect CSI carry over to the imperfect CSI case. We then use uncoded BER results to validate the pilot configurations introduced above and to justify focusing on the Gauss--Sinc filter before proceeding to the spectral efficiency analysis.\\
\begin{figure}[h]
    \centering
    
    \begin{minipage}{\columnwidth}
        \centering
        \includegraphics[width=\linewidth]{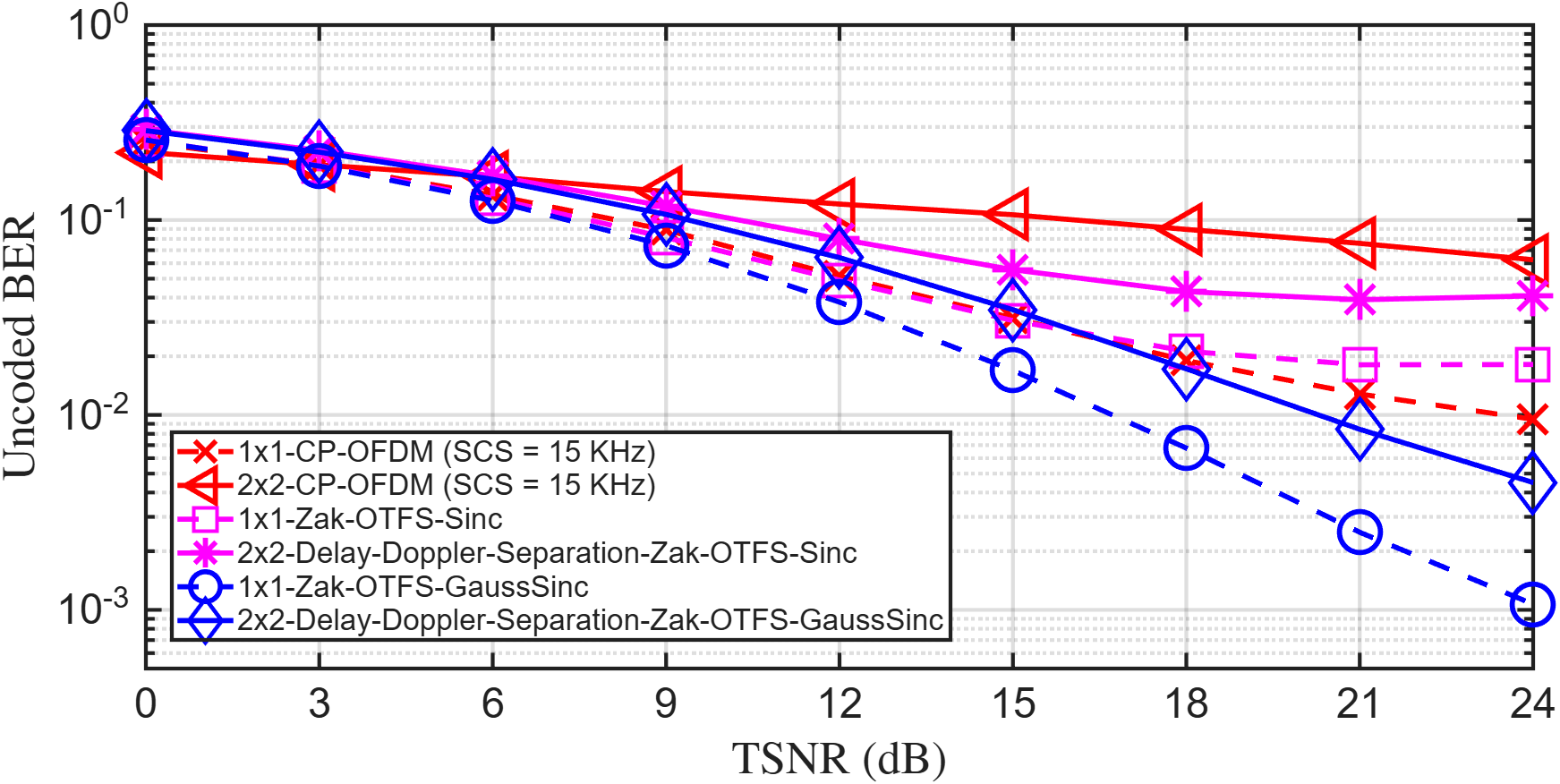}
        \caption*{(a) $\nu_{\max}=600\,\mathrm{Hz}$}
    \end{minipage}
    
    \vspace{0.3cm}
    
    \begin{minipage}{\columnwidth}
        \centering
        \includegraphics[width=\linewidth]{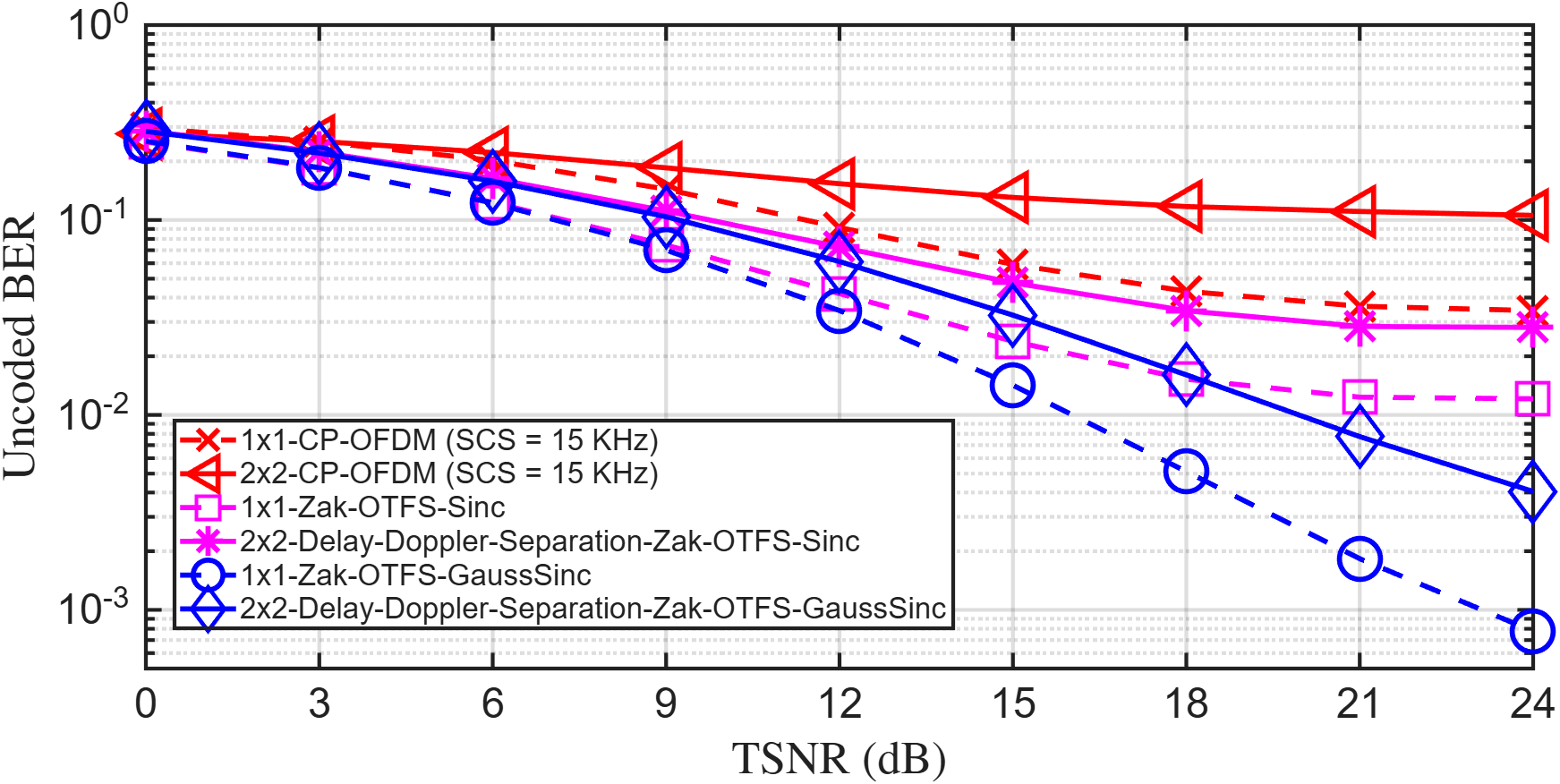}
        \caption*{(b) $\nu_{\max}=1200\,\mathrm{Hz}$}
    \end{minipage}
    
    \caption{Uncoded BER performance for Zak-OTFS and CP-OFDM under imperfect CSI with PDR $=-5~\rm dB$.}
    \label{fig:Uncoded_BER}
\end{figure}
\\
Fig~\ref{fig:Uncoded_BER} presents the BER performance under imperfect CSI for maximum Doppler spreads of $\nu_{\mathrm{max}} = 600$ and $1200,\mathrm{Hz}$, comparing CP-OFDM and Zak-OTFS. For CP-OFDM, a single pilot configuration is adopted, based on prior spectral efficiency optimization. For Zak-OTFS BER results, only the delay-Doppler separation has been investigated. It is observed that, with increasing TSNR, Zak-OTFS consistently outperforms CP-OFDM in both SISO and MIMO configurations across all Doppler scenarios. CP-OFDM exhibits performance saturation in high-Doppler regimes due to ICI, whereas Zak-OTFS avoids this limitation.

The choice of pulse-shaping filter plays a critical role. The sinc filter suffers from performance saturation due to poor localization, resulting in interference between data and pilot pulsones, thereby degrading channel estimation. On the other hand, the Gauss-sinc filter is more localized than the sinc filter resulting in better channel estimation and therefore improved performance compared to the sinc filter.

Based on these findings, only the Gauss-sinc filter is considered in the spectral-efficiency analysis presented in the subsequent subsections.

\subsection{Spectral Efficiency vs Doppler}
The Doppler resilience of SISO Zak-OTFS has been established in prior works~\cite{Zak_io}. Here, we demonstrate how this behavior extends to MIMO scenarios. Moreover, we confirm the expected fundamental crossover behavior: CP-OFDM performs better at low Doppler due to more efficient pilot configurations, while Zak-OTFS exhibits superior performance at higher Doppler owing to its inherent robustness.  Furthermore, we argue that this crossover point shifts to lower Doppler shift values as TSNR increases.

\begin{figure}[t]
    \centering

    \begin{minipage}{\columnwidth}
        \centering
        \includegraphics[width=\linewidth]{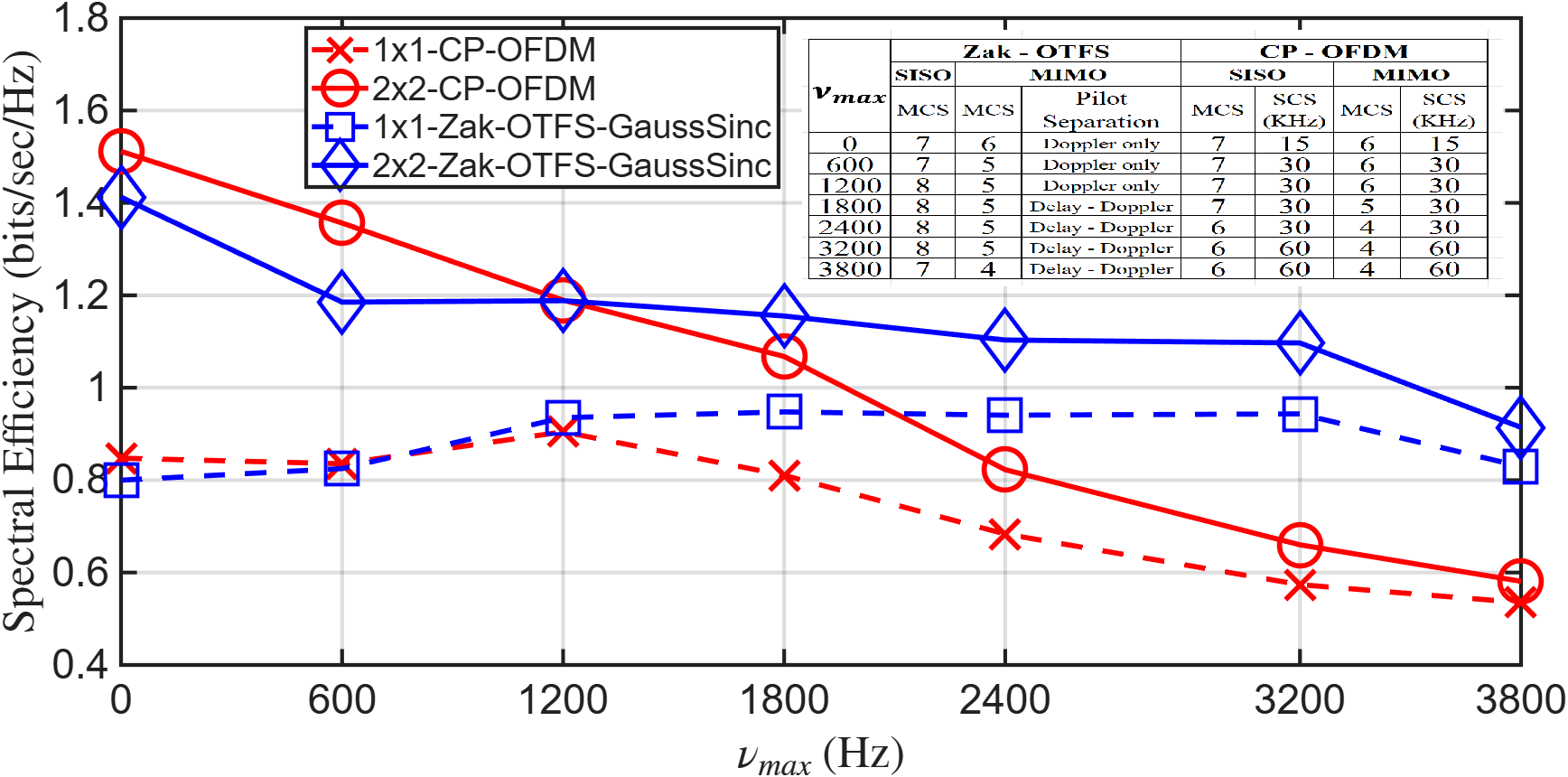}
    \end{minipage}

    \vspace{-2mm}

    \begin{minipage}{\columnwidth}
        \centering
        \includegraphics[width=\linewidth]{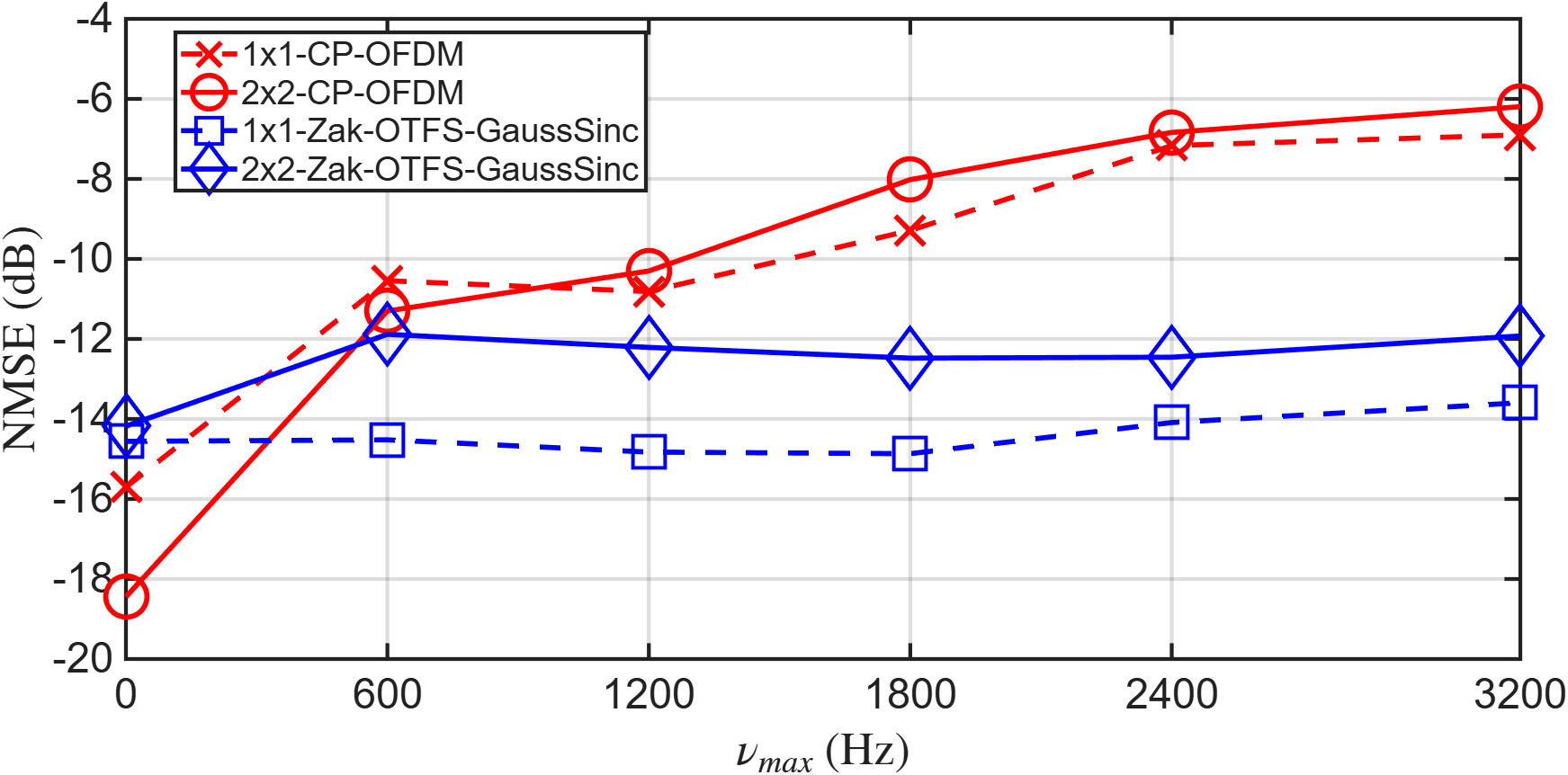}
    \end{minipage}

    \caption{Throughput and NMSE versus $\nu_{\rm max}$ for Zak-OTFS and CP-OFDM with TSNR $=12~\mathrm{dB}$ and PDR $=-5~\mathrm{dB}$.}
    \label{fig:SEvsnumax}
\end{figure}

Fig~\ref{fig:SEvsnumax} illustrates the spectral efficiency vs maximum Doppler shift of the two modulation schemes at the fixed $12\,\rm dB$ TSNR. In both cases, at very low Doppler, CP-OFDM outperforms Zak-OTFS while achieving the same MCS. This behavior is primarily due to more efficient pilot utilization in CP-OFDM. At low Doppler, the channel varies slowly over time, requiring fewer pilot symbols along the time dimension. In contrast, Zak-OTFS employs a fixed pilot overhead, resulting in less efficient pilot utilization in this regime.

In the SISO case, a slight increase in spectral efficiency is observed as the Doppler spread increases from very low to moderate values. This behavior can be attributed to the reduction in channel coherence time, which introduces temporal variations within a frame. As a result, symbols within a codeword experience partially independent channel realizations, enabling the exploitation of time diversity through channel coding. In particular, LDPC coding across time allows the decoder to effectively average the fading over multiple channel states, leading to improved reliability and reduced BLER.

More specifically, for SISO Zak-OTFS, in the range from $0$ to $600~\mathrm{Hz}$, the MCS index remains unchanged and the observed gain is primarily due to improved BLER. As the Doppler increases further from $600~\mathrm{Hz}$ to $1200~\mathrm{Hz}$, the improved reliability enables the selection of a higher MCS index, resulting in an additional increase in spectral efficiency. However, this overall gain remains modest.

This effect is not observed in the MIMO case. For both modulation schemes, the benefits of time diversity become less pronounced in MIMO systems, as spatial multiplexing dominates the performance and reduces the relative impact of temporal variations.

As the Doppler shift increases further, Zak-OTFS demonstrates significantly greater robustness, maintaining stable performance up to approximately $3200~\mathrm{Hz}$. Beyond this point, performance degradation is observed due to interference between the pilot pulsones transmitted from the two transmit antennas. In contrast, CP-OFDM experiences a pronounced performance loss with increasing Doppler, primarily due to ICI. The exact same behaviors are observed with respect to the channel estimation as well (see NMSE versus $\nu_{max}$ in Fig \ref{fig:SEvsnumax}). Overall, CP-OFDM is more efficient at very low Doppler due to lower pilot overhead, whereas Zak-OTFS provides superior robustness and sustained performance at moderate-to-high Doppler, with time diversity gains in SISO and ICI resilience dominating at higher mobility.
\subsection{Spectral Efficiency vs TSNR}
\begin{figure}[t]
    \centering

    \begin{minipage}{\columnwidth}
        \centering
        \includegraphics[width=\linewidth]{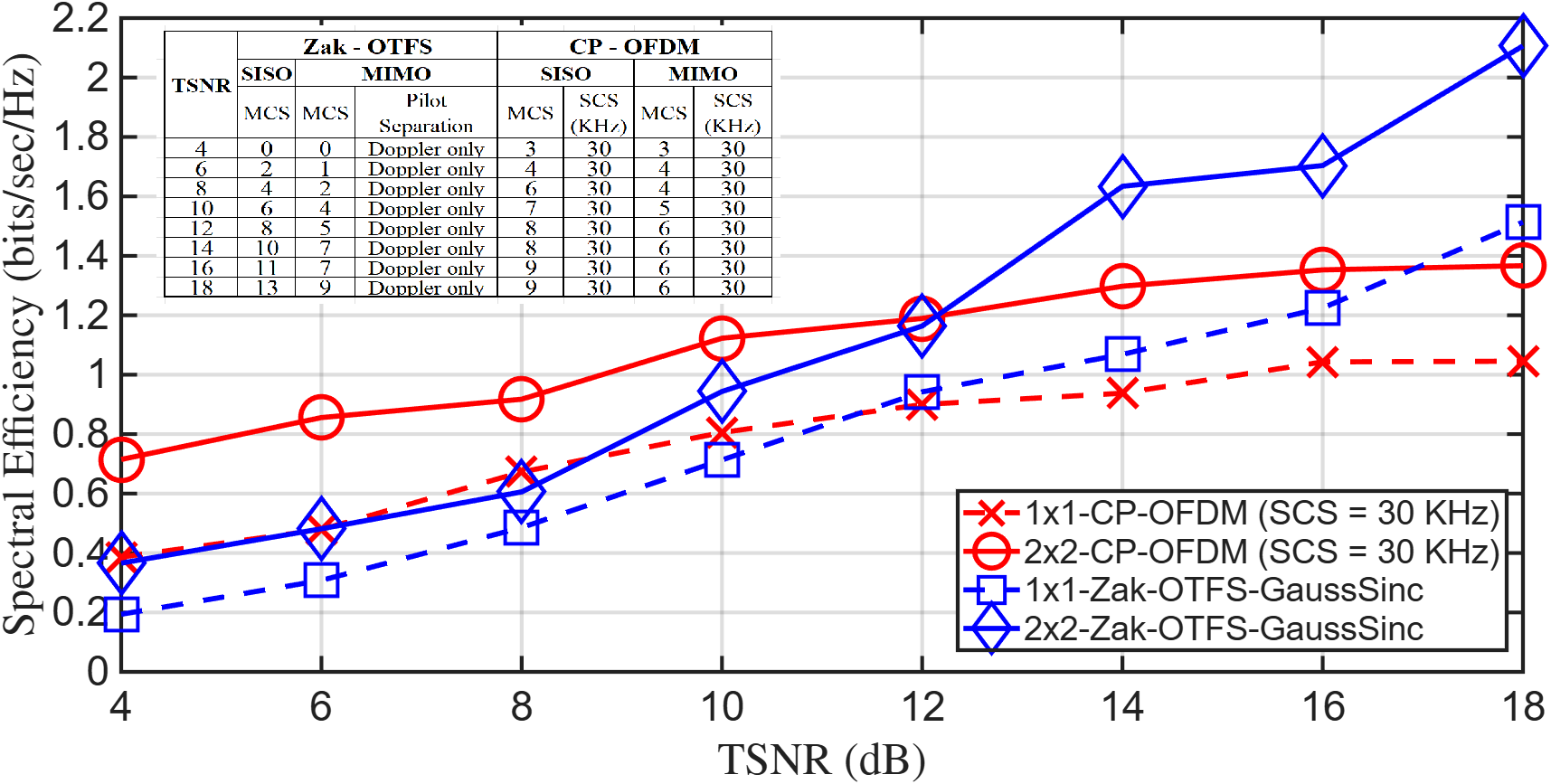}
    \end{minipage}

    \vspace{-2mm}

    \begin{minipage}{\columnwidth}
        \centering
        \includegraphics[width=\linewidth]{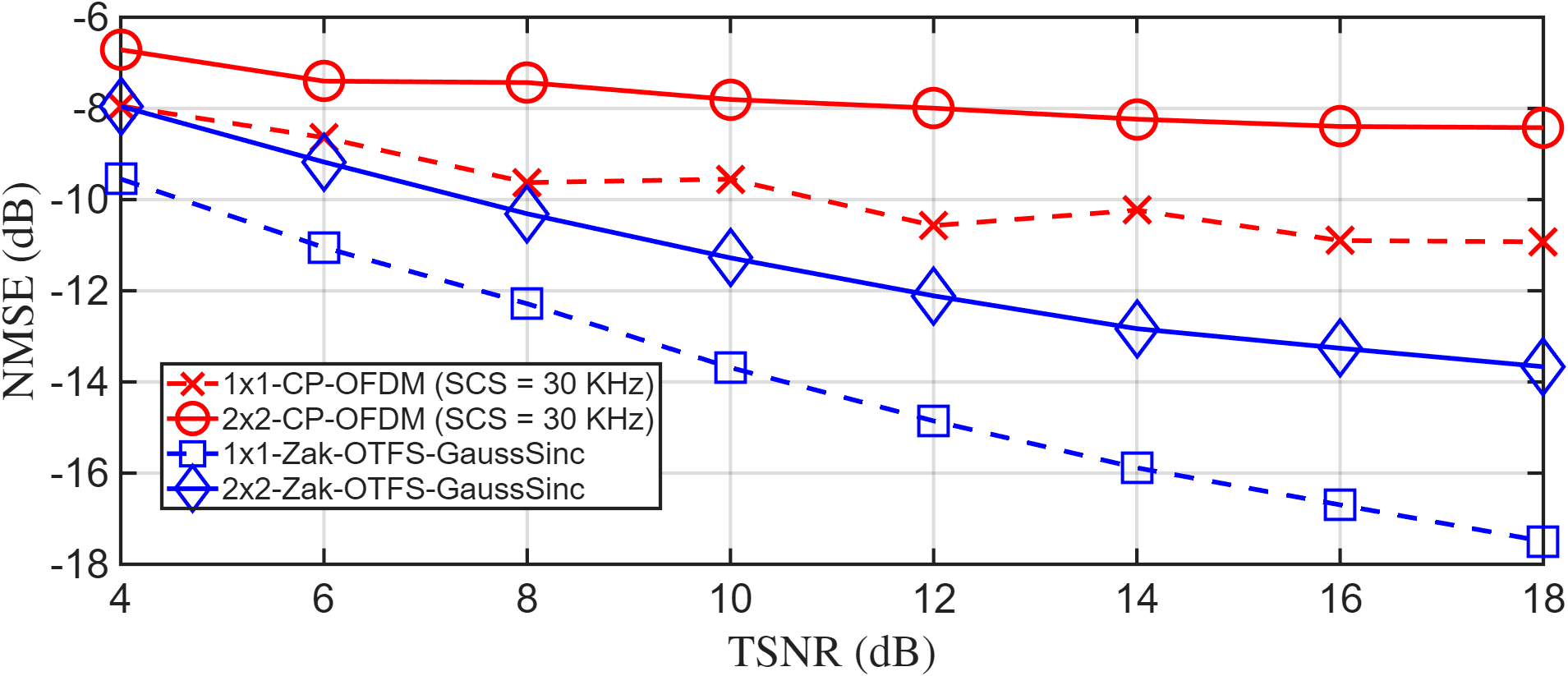}
    \end{minipage}

    \caption{Throughput and NMSE versus TSNR for Zak-OTFS and CP-OFDM with $\nu_{\rm max}=1200~\mathrm{Hz}$ and PDR $=-5~\mathrm{dB}$.}
    \label{fig:SEvsTSNR}
\end{figure}

Next, we demonstrate a similar crossover behavior with respect to the TSNR. At low TSNR, CP-OFDM performance is not limited by ICI as ICI is dominated by noise. At the same time, Zak-OTFS channel estimation error is high due to low pilot to noise ratio (as PDR is fixed) resulting in lower MCS for Zak-OTFS when compared to CP-OFDM (see the MCS table inside Fig~\ref{fig:SEvsTSNR}). At high TSNR, however, CP-OFDM performance saturates due to ICI, whereas Zak-OTFS continues to improve without exhibiting such a limitation. Furthermore, we argue that this crossover point shifts to lower TSNR values as the Doppler spread increases. This is because for higher Doppler, the ICI is higher and therefore noise dominates ICI at a lower TSNR. In other words, at higher Doppler CP-OFDM performance gets limited by ICI at a lower TSNR.

Fig~\ref{fig:SEvsTSNR} illustrates the performance of Zak-OTFS and CP-OFDM as a function of TSNR. A crossover point between the two modulation schemes is observed.

 At high TSNR, the performance of CP-OFDM, in both SISO and MIMO configurations, is ultimately limited by fading and ICI, leading to a saturation behavior at high TSNR. In contrast, Zak-OTFS inherently accounts for the doubly dispersive nature of the channel by representing the signal in the delay-Doppler domain and jointly equalizing the effects of delay and Doppler, allowing the throughput to continue increasing with TSNR. It has been observed that with increasing Doppler shift, the crossover point moves to the left, i.e. Zak-OTFS outperforms CP-OFDM at lower TSNR values.
\subsection{Spectral Efficiency vs PDR}
Finally, we justify the choice of the PDR value used throughout the paper and show that Zak-OTFS is significantly more sensitive to variations in PDR.
\begin{figure}[h!]
       \centering
        \includegraphics[width=\linewidth]{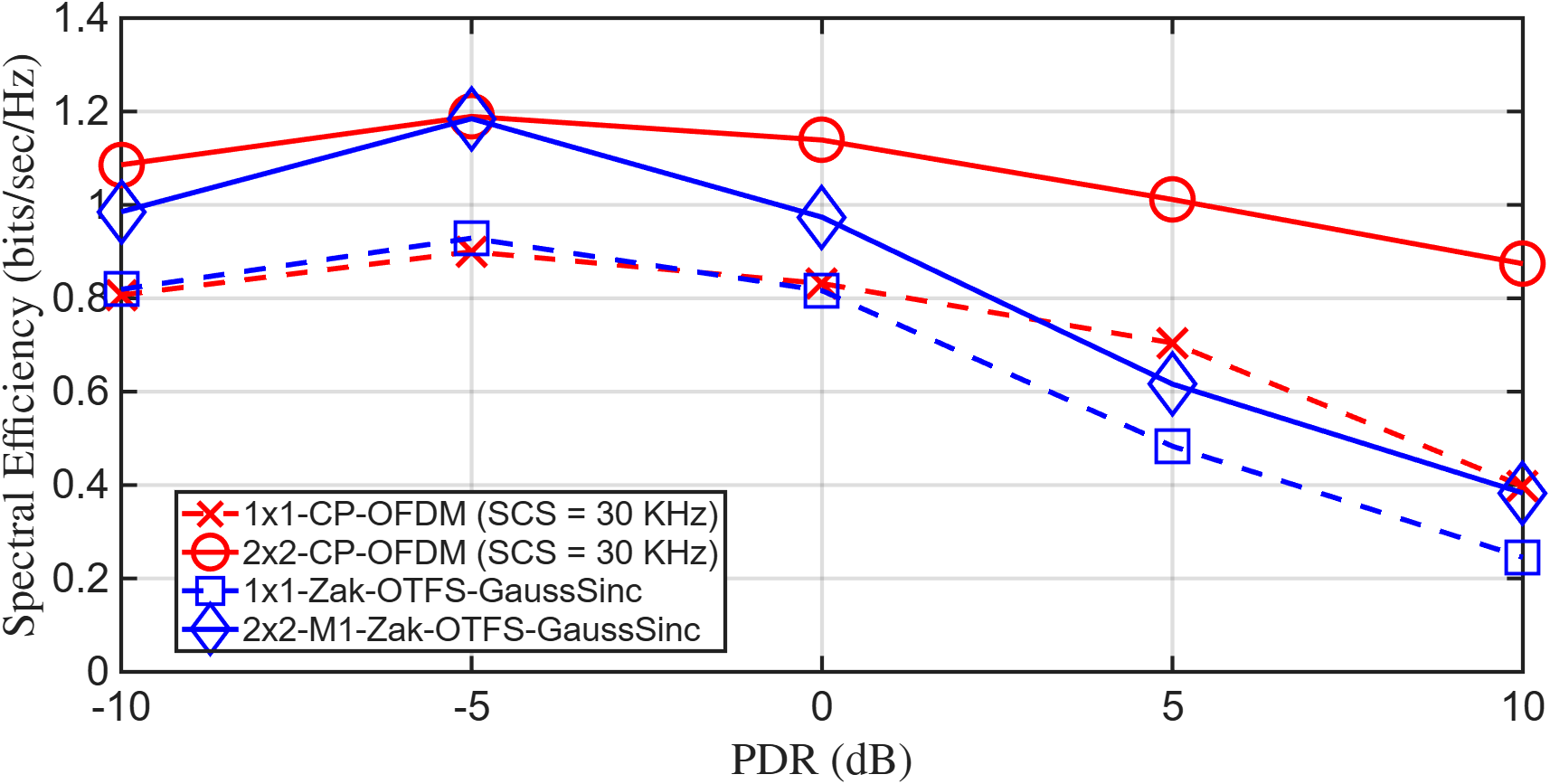}

    \caption{Throughput vs $\rm PDR$  for Zak-OTFS and CP-OFDM with TSNR$=12~\rm dB$ and $\nu_{\rm max} = 1200~\rm Hz$.}
    \label{fig:SEvsPDR}
\end{figure}

Fig~\ref{fig:SEvsPDR} illustrates the achieved spectral efficiency as a function of PDR. It can be observed that the optimal PDR for both modulation schemes, under both SISO and MIMO configurations, occurs at $-5$ $\mathrm{dB}$. Notice that the operating points corresponding to TSNR and $\nu_{\rm max}$ lie very close to the crossover region, as shown in Figs ~\ref{fig:SEvsnumax} and~\ref{fig:SEvsTSNR}.
Furthermore, reducing the data power has a more pronounced impact on Zak-OTFS compared to OFDM. In both SISO and MIMO scenarios, the spectral efficiency of Zak-OTFS degrades more rapidly as PDR increases. For small PDR, pilot power to noise power ratio is small due to which channel estimation suffers resulting in lower SE. With increasing PDR, channel estimation improves resulting in improvement in SE. However, with further increase in PDR, the interference of pilot carriers to data carriers (both for CP-OFDM and Zak-OTFS) increases which then decreases SE. In CP-OFDM, the pilot power is spread over several pilot carriers and therefore the power level of each pilot carrier is much smaller compared to the single pilot carrier in Zak-OTFS. The peaky single pilot carrier in Zak-OTFS creates stonger interference for data carriers on neighbouring DD taps as compared to the interference from CP-OFDM pilot carriers to CP-OFDM data carriers. This explains why Zak-OTFS SE is more sensitive to PDR as compared to CP-OFDM SE.


\section{Conclusion}
This paper introduces a novel extension of the Zak-OTFS framework to MIMO systems. First, we develop a unified I/O relation for MIMO Zak-OTFS based on a physical channel representation and antenna geometry. Building on this formulation, we propose a scalable channel estimation method that enables the MIMO extension without incurring a linear increase in pilot overhead with increasing number of transmit antennas. Finally, we provide a comprehensive performance evaluation against MIMO CP-OFDM in terms of bit error rate and spectral efficiency across a wide range of TSNR and Doppler conditions.

The results show that the fundamental advantages and trade-offs observed in SISO systems largely carry over to the MIMO setting. In particular, a crossover behavior is observed between the two modulation schemes in term of spectral efficiency when sweeping over different Doppler shifts or different TSNR values. CP-OFDM tries to avoids ISI and ICI by adapting its numerology, which becomes increasingly inefficient in highly doubly dispersive channels. In contrast, Zak-OTFS explicitly models and equalizes these effects in the delay-Doppler domain, but its performance is limited by receiver equalization complexity.

These findings highlight a key design trade-off between robustness to channel dispersion and receiver complexity, suggesting that Zak-OTFS is a suitable candidate for high-mobility MIMO scenarios where Doppler effects are significant.

\bibliographystyle{IEEEtran}
\bibliography{bib}

\end{document}